\PassOptionsToPackage{table}{xcolor}
\documentclass[10pt,journal,compsoc]{IEEEtran}

\usepackage{libertine}
\usepackage[T1]{fontenc}
\usepackage[utf8]{inputenc}

\newcommand{\subparagraph}{} 
\usepackage[compact]{titlesec}

\usepackage{blindtext, graphicx}
\usepackage{amsfonts}

\usepackage{amsmath, amsthm, amssymb, nccmath}
\usepackage{times}
\usepackage{url}
\usepackage{verbatim}
\usepackage{array}
\usepackage{mathrsfs}
\usepackage{enumitem}
\usepackage{floatflt}
\usepackage{mathtools}
\DeclarePairedDelimiter{\ceil}{\lceil}{\rceil}
\DeclareMathSizes{10}{9}{7}{5} 

\usepackage[ruled, lined]{algorithm2e}
\SetAlFnt{\small}

\SetCommentSty{mycommfont}

\usepackage{amsfonts}
\usepackage{bbm}

\usepackage{color}
\usepackage{graphics}
\usepackage{graphicx,epsfig}
\usepackage{lipsum,graphicx,subcaption}
\graphicspath{{figures/}}
\usepackage{caption}
\usepackage{subcaption}

\usepackage{cleveref}
\usepackage{wrapfig}
\usepackage{url}
\DeclareMathOperator*{\argmax}{argmax}

\usepackage{tabularx}
\usepackage{comment}
\usepackage{multirow}
\usepackage{makecell}

\usepackage{footnote}
\usepackage{tablefootnote}

\usepackage[table,xcdraw]{xcolor}

\newtheorem{theorem}{Theorem}

\newtheorem{example}[theorem]{Example}

\newcommand{\note}[1]{{\color{black}{#1}}} 
 
\newcommand{\anothernote}[1]{{\color{blue}{#1}}} 

\newcommand{\rood}[1]{\textbf{\textcolor{red}{(#1)}}} 

\newcommand{\abbrev}{\ttfamily{RFEye}}
\newcolumntype{P}[1]{>{\centering\arraybackslash}p{#1}}
\newcolumntype{x}[1]{>{\centering\arraybackslash\hspace{0pt}}p{#1}}

\ifCLASSOPTIONcompsoc
  \usepackage[nocompress]{cite}
\else
  \usepackage{cite}
\fi

\begin{document}
\def\eg{\mbox{\em e.g.}, }



\title{{\abbrev} in the Sky}





\author{Maqsood~Ahamed~Abdul~Careem,~\IEEEmembership{Student Member,~IEEE,}
        Jorge Gomez,~\IEEEmembership{Student Member,~IEEE,}\\
        Dola Saha,~\IEEEmembership{Member,~IEEE,}
        and~Aveek~Dutta,~\IEEEmembership{Member,~IEEE}
\IEEEcompsocitemizethanks{\IEEEcompsocthanksitem The authors are with the Department
of Electrical and Computer Engineering, University at Albany SUNY, NY,
12222. 
(e-mail: mabdulcareem@albany.edu; jgomez4@albany.edu; dsaha@albany.edu; adutta@albany.edu).}}

\markboth{This article has been accepted for publication in 
IEEE Transactions on Mobile Computing}{} %
%

\IEEEtitleabstractindextext{%
\begin{abstract}
We introduce {\abbrev}, a generalized technique to locate signals independent of the waveform, using a single Unmanned Aerial Vehicle (UAV) equipped with only one omnidirectional antenna. This is achieved by acquiring signals from uncoordinated positions within a sphere of 1-meter radius at two nearby locations and formulating an asynchronous, distributed receiver beamforming at the UAV to compute the Direction of Arrival (DoA) from the unknown transmitter. The proposed method includes four steps: 1) Blind detection and extraction of unique signature in the signal to be localized, 2) Asynchronous signal acquisition and conditioning, 3) DoA calculation by creating a virtual distributed antenna array at UAV and 4) Obtaining position fix of emitter using DoA from two locations. These steps are analyzed for various sources of error, computational complexity and compared with widely used signal subspace-based DoA estimation algorithms. {\abbrev} is implemented using an Intel-Aero UAV, equipped with a USRP B205 software-defined radio to acquire signals from a ground emitter. Practical outdoor experiments show that {\abbrev} achieves a median accuracy of 1.03m in 2D and 2.5m in 3D for Wi-Fi, and 1.15m in 2D and 2.7m in 3D for LoRa (\note{Long Range}) waveforms, and is robust to external factors like wind and UAV position errors.
\end{abstract}

}

\maketitle

\IEEEdisplaynontitleabstractindextext
\IEEEpeerreviewmaketitle

\section{Introduction}
\label{sec:intro}

Locating wireless signals in outdoor environment is necessary for various applications like locating rogue emitters for enforcing spectrum policies, pin-pointing targets in search and rescue operations and tracking the movement of agents in electronic warfare. Acquiring such signals is the first step, which often requires patrolling large distances over varied terrain and can be limited if terrestrial methods, like autonomous vehicles, crowdsourcing or fixed receiver arrays are used. Also, the signal acquired by terrestrial means can be severely attenuated or impaired due to Non-Line-of-Sight (NLOS), shadowing and other unknown factors. 
Signal acquisition using UAVs can overcome such limitations by flying to advantageous locations but are constrained on resources like power, RF front-end, antenna, processing and storage. {\abbrev} is the first of its kind, that computes accurate location of a wireless emitter using only \textit{one} UAV, equipped with only \textit{one} antenna, acquiring no more than $20$ packets in only \textit{two} locations, separated by few meters without any precise maneuvering of the UAV. 
\begin{figure}
     \centering
     \includegraphics[width=0.99\linewidth]{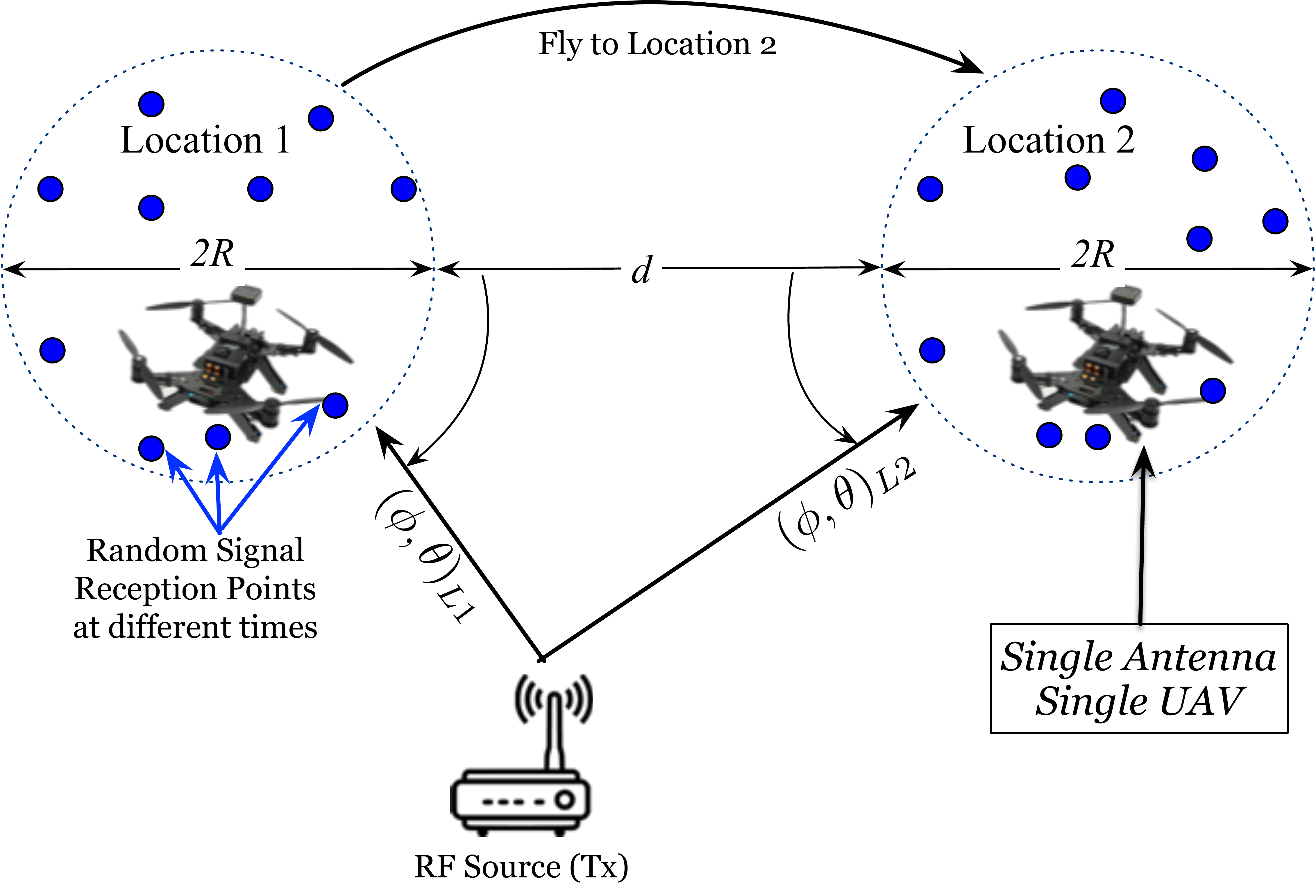}
     \caption{{\abbrev} in action: One UAV equipped with only one antenna hovers at two locations to acquire signals from a remote emitter to localize it.}
     \label{fig:system}
\end{figure}

Figure~\ref{fig:system} shows the operation of {\abbrev}. The UAV flies to an initial location of interest (Location 1) and {hovers} within a sphere of radius 1 meter, collecting signals using one omnidirectional antenna at various positions within the sphere. Then, the UAV moves to Location 2, a few meters away, and repeats the same procedure. These sets of signals acquired in each location are {fused} using a novel alignment technique 
while preserving the phase offsets between the signals acquired at different positions. The novelty of {\abbrev} is in leveraging the normal movement of the UAV to acquire wireless signals without prior knowledge of the signal type or its transmit power 
and then align those in time to obtain the benefits of synchronous reception from spatially distributed antenna elements. All of this is achieved without the overhead of synchronous reception from multiple receiver front-ends. The temporal alignment is crucial to beamform using random acquisition points and to calculate the DoA ($\phi{=}$Azimuth, $\theta{=}$Elevation) from the emitter by searching the beamspace for the most likely direction of the signal. Further, by combining this information from two locations, as shown in Figure~\ref{fig:system}, an accurate fix is obtained using trigonometric identities. 

{\abbrev} is designed to be highly portable to any UAV where reducing power consumption and payload is a priority. Instead of spending prohibitive amount of power to position the UAV on specific coordinates~\cite{drone_control_16}, {\abbrev} embraces the normal movement of the UAV to acquire signals, including displacement due to wind, thus emulating a distributed antenna array. Wireless signals, especially waveforms used for communication purposes, often contain repetitive patterns (e.g., cyclostationary features embedded in preambles of physical layer packets) that are repeated periodically for error-free communication. Careful detection and alignment of such repetitive patterns embedded in these signals is used to extend the mathematical foundation of distributed synchronous arrays to design robust DoA algorithms in {\abbrev} for uncoordinated and asynchronous arrays. 
Therefore, the contributions of this work are as follows:


\newcounter{contrib}

\stepcounter{contrib}
\noindent
 \textbf{\arabic{contrib}) Resource saving signal acquisition using single UAV} -- {\abbrev} relies on a single UAV equipped with one antenna, with no precise control for positioning of the UAV. This greatly simplifies the payload and reduces the power requirement for the carrier. 
Furthermore, we design a computationally lightweight algorithm for on-board signal processing.

\stepcounter{contrib}
\noindent
\textbf{\arabic{contrib}) Waveform independent signal acquisition} -- By utilizing repetitive patterns embedded at the start of packet, {\abbrev} implements a novel blind signature detection algorithm that removes any dependency on the emitter waveform. 

\stepcounter{contrib}
\noindent
\textbf{\arabic{contrib}) Asynchronous distributed receiver beamforming} -- 
We emulate a distributed virtual phased array receiver that combines the measured positions of the UAV 
and time aligned features to compute the beampattern in each of the $360^\circ$ along the azimuth and elevation. The direction that provides the maximum accumulated power in the beampattern, yields the DoA.

\stepcounter{contrib}
\noindent
\textbf{\arabic{contrib}) Error minimization using clustering} -- 
We explore the beamspace by combining a small subset of the captured signal each time, yielding a larger number of candidate DoAs. We use clustering to choose the median of the dominant cluster as the DoA. Furthermore, we analyze the sources of errors in {\abbrev} and compare that with subspace-based DoA algorithm\note{~\cite{music_86}}. 

\stepcounter{contrib}
\noindent
\textbf{\arabic{contrib}) \note{Hardware platform using COTS (Commercial off-the-shelf) devices}} -- {\abbrev} is a fully deployable platform, equipped with all the software components necessary to acquire signals in the field. The hardware architecture includes Intel Aero UAV, USRP B205-mini software-defined radio, Raspberry Pi and RTK-GPS modules. 
The research has been validated with extensive outdoor experiments using this platform.

\section{Related Work}

\begin{table*} 
\caption{Comparison with state-of-the-art RF localization systems}
\centering
\scriptsize
\begin{tabular}{ |l|c|c|c|c|c|c|l|}
\hline
\makecell[l]{\textbf{System}}                      & \makecell[l]{\textbf{Metric}}         &  \makecell[c]{\textbf{Method}}        & \makecell[c]{\textbf{Accuracy} \\\textbf{(Median)}}& \makecell[l]{\textbf{Number of} \\\textbf{Receivers}}       & \makecell[l]{\textbf{Number of} \\\textbf{Antennas}}        & \makecell[c]{\textbf{Bandwidth}}         & \makecell[c]{\textbf{Generality}}        \\
\hline
ArrayTrack \cite{ArrayTrack}& AOA               &  MUSIC              & \makecell[c]{0.23m (2D)\\(6 APs, 8 antenna)}   & \makecell[c]{3\\ (min) }                        & \makecell[c]{4\\ (min) }                         & 40MHz                & Limited to Wi-Fi  \\
\hline
ToneTrack \cite{ToneTrack}  & TDOA               &  MUSIC              & \makecell[c]{0.90m \\(4 APs, 3 antenna)}    & \makecell[c]{3\\ (min) }                        & \makecell[c]{3\\ (min) }                         & 
\makecell[c]{3 of 20MHz \\channels}                & Limited to Wi-Fi   \\
\hline
SpotFi \cite{Spotfi}        & \makecell[c]{AOA, \\ ToF}        &  MUSIC              & \makecell[c]{0.4m (2D) \\(6 APs, 3 antenna)}   & \makecell[c]{3\\ (min)}                       & 3                         & 40MHz          & Limited to Wi-Fi   \\
\hline
Ubicarse \cite{Ubicarse}    & AOA               &  SAR                & \makecell[c]{0.39m \\(5 APs)}    & \makecell[c]{3\\ (min)}                        & 2                         & 
\makecell[c]{20/40MHz}                   & \makecell[l]{Limited to Wi-Fi}   \\
\hline
RFly \cite{RFly}    & AOA               &  SAR                & 0.19m (2D)    & \makecell[c]{1 Reader, \\ 1 Relay} & \makecell[c]{1 on Reader, \\4 on Relay}                          &                   & \makecell[l]{Limited to RFID}\\
\hline
Chronos \cite{chronos_16}   & ToF               &  ToF Ranging        & 0.65m (2D)    & 1                         & 3                         & \makecell[c]{Multiple 20MHz \\channels}                   & Limited to Wi-Fi   \\
\hline
SAIL \cite{Sail}            & ToF               &  \makecell[c]{ToF Ranging,\\ Deadreckoning} & 2.3m (2D)    & 1               & 3                         & 40MHz                  & \makecell[l]{Limited to Wi-Fi}   \\
\hline
System in \cite{RSSI}       & RSSI              & Trilateration       & \makecell[c]{0.83m \\(for distance$\leq$2.5m)}  & 3     & 1                         & \makecell[c]{Any \\Bandwidth}                & \makecell[l]{Known Tx power\\ Any waveform} \\
\hline
\rowcolor[HTML]{FFFFBF}
\textbf{{\abbrev}}         & \textbf{AOA}               &  \textbf{Distributed} & \textbf{Wi-Fi: 1.03m (2D)}& \textbf{1}                         & \textbf{1}       & \textbf{Any} & \textbf{Any waveform}\\
\rowcolor[HTML]{FFFFBF}& & \textbf{Receiver}&\quad\quad\quad\textbf{2.50m (3D)} & & &\textbf{Bandwidth} &\textbf{with repetitive}\\ 
\rowcolor[HTML]{FFFFBF}& & \textbf{Beamforming}&\textbf{LoRa: }\textbf{1.15m (2D)} & & & &\textbf{pattern}\\
\rowcolor[HTML]{FFFFBF}& & &\quad\quad\quad\textbf{2.70m (3D)} & & & &\\
\hline
\end{tabular}
\label{tab:comparison}
\end{table*}

Precise localization of an emitter in indoor or outdoor settings has been studied for decades~\cite{loc_book_11}. In this section, we discuss the literature that is most relevant to this work. A broad comparison of {\abbrev} to state-of-the-art RF localization technique is shown in Table \ref{tab:comparison}. The key difference of {\abbrev} is that, unlike other methods, it is able to localize any RF source (that transmits signals with repetitive patterns in it), by using a single receiver with only a single antenna, and achieves reasonable 
localization accuracy.

\noindent
\textbf{Distinction between SAR and distributed beamforming}:
Synthetic Aperture Radars (SAR) use coherent detection of signals reflected from a target to emulate an antenna aperture to generate 
remote sensing imagery or locate the target. In contrast, {\abbrev} has the advantage of localizing an unknown emitter itself using the signals captured at multiple locations using a single UAV. Localization based on SAR also leverage the movement of a device acting as a receiver to localize itself using signals captured from multiple transmitters \cite{Ubicarse}. Whereas in {\abbrev}, a single mobile receiver can locate a target emitter, 
which eliminates the need for multiple transmitters. 

\noindent
\textbf{Signal localization from UAVs}: {\abbrev} is one of many applications enabled by UAVs~\cite{uav_tutorial,HiperV}. 
TrackIO~\cite{TrackIO} uses an outdoor UAV to localize mobile indoor nodes using Ultra Wideband (UWB) signals, trilateration and ranging protocols. Authors in SensorFly~\cite{SensorFly} propose an indoor aerial sensor network of small UAVs that can self-locate using anchor nodes, which is an added infrastructure. 
In \cite{SkyRAN}, the transmitter's location is derived by using UAV’s GPS and Time of Flight (ToF) data using multilateration.  
Alternately, authors in \cite{Garza} use a UAV mounted 3D antenna array to maximize directivity, while in~\cite{Liu}, a single portable helicopter based localization system is presented using the Received Signal Strength Indicator (RSSI) for localization, which fundamentally limits the accuracy of the derived location. In~\cite{drone_loc_18}, authors use directional antenna to find the direction of arrival of the source. However, directional antennas limit visibility of the radio environment often leading to blind spots and misdetection of unknown signals. Authors in~\cite{network_uav} use a network of UAVs to localize GPS jammers similar to military applications, where multiple UAVs~\cite{uav_localization_15, uav_locate_16}   
are used to scan in different channels to avoid jammers. Evidently, all of these methods require additional resources to be deployed for the system to work. Simultaneous localization of UAV and RF sources is proposed in ~\cite{slus_14}, that also include terrain information~\cite{terrain_13} to improve accuracy. UAV wireless channel models are shown to be complicated and are extensively studied~\cite{Khuwaja}. However, we obviate any dependence on such deterministic or empirical models of the channel, by employing the geometry of distributed beamforming. 
Finally, RFly \cite{RFly} uses an RFID reader and RFID relay on a UAV to locate RFID tags by emulating an antenna array using SAR. In contrast {\abbrev} localizes any source that emits a waveform with a repetitive pattern, which is widely known to be a common feature in communication signals.

\noindent
\textbf{Indoor signal localization:}
Almost all indoor localization techniques, including ArrayTrack~\cite{ArrayTrack}, ToneTrack~\cite{ToneTrack}, PinPoint~\cite{PinPoint}, Spotfi~\cite{Spotfi}, primarily employ commonly known signal subspace-based methods (like MUltiple SIgnal Classification (MUSIC)~\cite{music_86}) either to derive Angle of Arrival (AOA) or Time Difference of Arrival (TDOA) and also require multiple antennas and wide bandwidths for accuracy. 
However, accuracy of DoA estimation using subspace methods are shown to severely degrade in the presence of array uncertainties such as errors in element position and synchronization~\cite{MUSIC_Sensitivity,loc_book_11}. 
Chronos~\cite{chronos_16} uses ToF data and relies on a MIMO access point (AP) to localize a transmitter associated with it. 
Ubicarse~\cite{Ubicarse} relies on the movement of a mobile, MIMO receiver (with at least two antennas) to emulate an antenna array and uses SAR techniques on the signals emitted from multiple transmitters to localize the receiver. 
This requires additional hardware (multiple antenna) for locating a single device, which may not be available in outdoor environments and not suited for resource constrained devices. Moreover, it relies on the channel states measured by a Wi-Fi network card that cannot be used to localize other waveforms.
In contrast, {\abbrev} leverages only one antenna on a single receiver and uses receiver distributed beamforming on the repetitive patterns extracted from signals captured at various positions, to localize a remote RF transmitter. 

\noindent
\textbf{Outdoor signal localization:}
Literature on outdoor localization predominantly relies on GPS based self-positioning, which can help a source to locate itself within ${\approx} 3{-}5$m 2D location estimates~\cite{faa_gps_2016}. In contrast, if that source wants to be located by other services, it can transmit its GPS location. This is only possible if a) the GPS receiver is available on the source consuming additional power, b) the source is not in a GPS-denied environment, and c) the source broadcasts its GPS coordinates. 
However, there are many applications where these assumptions may not hold. For example: 
1) In search and rescue missions, the rescuee to be tracked can be located in GPS-denied environments (\eg debris, foliage, underground, etc.). Also, backpackers and hikers prefer gear that are wearable, energy efficient and long-range, which are reliable in variety of terrain and infrastructure~\cite{senstracking,semtechSmartHealth}, over more expensive, bulky and power-hungry GPS devices. 
2) In locating rogue emitters for spectrum policy enforcement and in electronic warfare, the devices will not broadcast their GPS coordinates to preserve privacy and security.  
In such cases, {\abbrev} accurately locates any unknown remote source by capturing its transmitted signals from an UAV without prior knowledge of the signal structure and is robust to a variety of terrain and infrastructure. 

\note{
\noindent
\textbf{High frequency and wideband techniques:}
Prior work have used other frequencies and wider bands to locate sources. For example, \cite{Bernhard} uses UWB to locate mobile tags using a single anchor along with a crude floor plan while \cite{Pefkianakis} provides 3D indoor location by exploiting the small wavelength and directional communication of mm-wave networks. Authors in~\cite{small_devices_18} introduce a multi-band backscatter technique to locate low power devices in non-line-of-sight scenarios and visible light is used 
in~\cite{rainbowlight_18}. 
However, many of these approaches are not feasible for localization of sources located at longer distances using a radio resource constrained receiver, which is why {\abbrev} has been designed to operate in sub-6GHz bands with only 5 to 20MHz signal bandwidth.
} 

\section{Background} 
\label{sec:Distributed}
\begin{figure}[h]
\vspace{-10pt}   
        \centering
		\includegraphics[width=.7\linewidth]{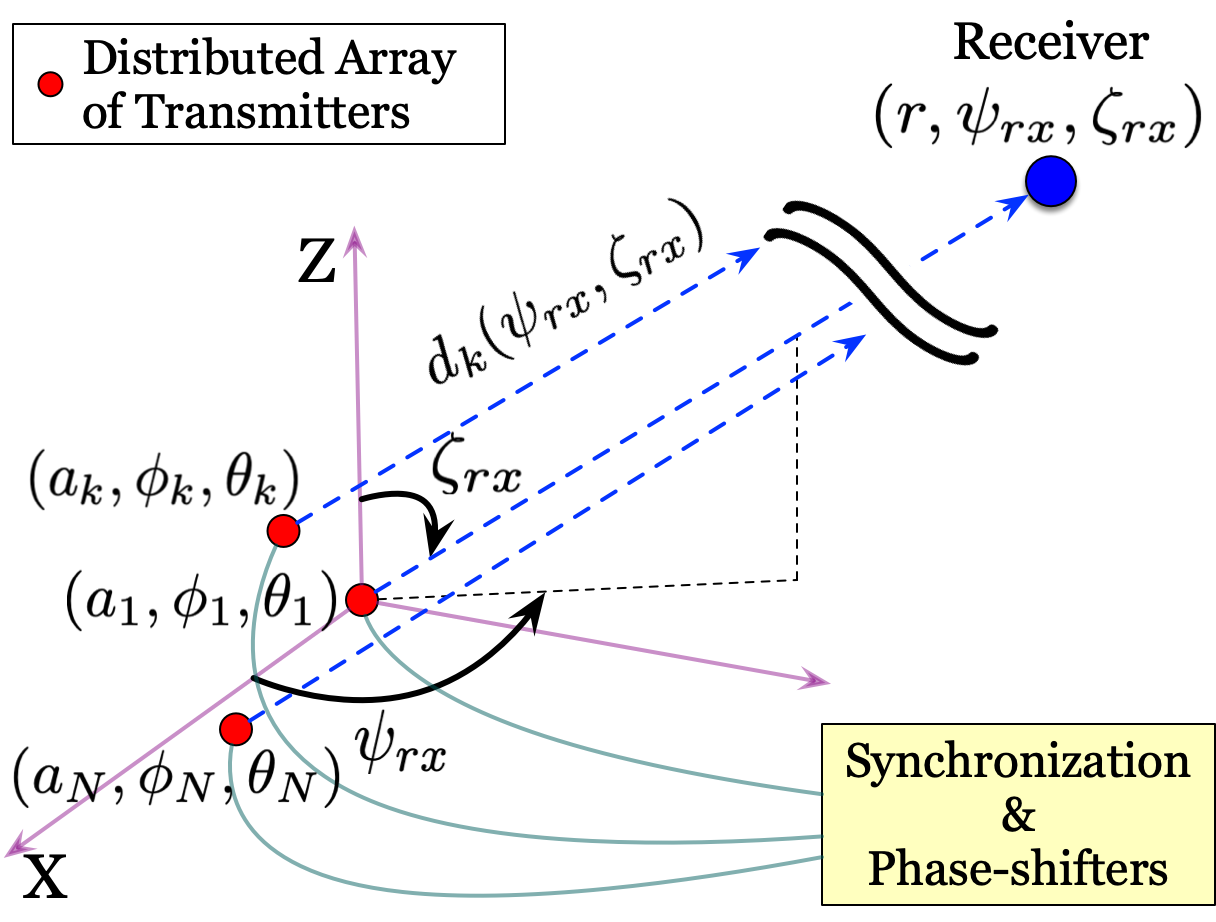}
        \caption{Distributed transmitter beamforming}  
        \label{fig:geometry}
    \vspace{-10pt}   
\end{figure}
Distributed beamforming~\cite{DistributedBF,dist_beam_07} has been applied for transmitter beamforming by a cluster of distributed transmitters to a distant base station by emulating the properties of a synchronous antenna array. This requires precise timing and carrier frequency synchronization for collaborative transmissions. 
Figure~\ref{fig:geometry} shows the geometry of a receiver (in blue) in far field and the synchronous transmitters (in red) that are randomly distributed in a three dimensional space. Since, the receiver is in far field, the electromagnetic waves emerge from the transmitter array in parallel, which is also the direction of arrival of the signal at the receiver.
The position of the $k^{th}$ transmitter in spherical coordinates is given by $(a_{k}, \phi_{k}, \theta_{k})$ and the first node located at $(a_{1}, \phi_{1}, \theta_{1})$ is taken as the origin, without loss of generality.
The location of the remote receiver is given by $\left(r, \psi_{rx}, \zeta_{rx} \right)$ 
and the angles, $\psi_{rx} \in[-\pi, \pi]$ and $\zeta_{rx} \in[-\pi, \pi]$ denote the azimuth and elevation of the receiver respectively. 
Assuming perfect synchrony among the transmitters to eliminate any frequency offset or phase jitter, 
the Euclidean distance between the $k^{th}$ transmitter at $(a_{k}, \phi_{k}, \theta_{k})$ and any point at $(r, \psi, \zeta)$ is given by (\ref{eq:euclid}), 
\begin{align}
d_k(\psi, \zeta)&{=}[r^{2}{+}a_k^{2}{-}2 r a_{k} (\sin \psi\sin \phi_k  \cos \left(\zeta{-}\theta_k\right) \nonumber\\
&\quad\quad\quad\quad{+}\cos \psi \cos \phi_k)]^{\frac{1}{2}} \nonumber\\
&\approx r - a_{k} \left(\sin \psi\sin \phi_k  \cos \left(\zeta-\theta_k\right)+\cos \psi \cos \phi_k\right)
\label{eq:euclid}
\end{align}
where $r \gg a_{k}$ in the far-field region.

Therefore, the phase of the incident signal at $(r, \psi, \zeta)$ due to the transmission from the $k^{th}$ transmitter is given by the complex exponential, $e^{j \frac{2 \pi}{\lambda} d_{k}\left(\psi, \zeta\right)}$. 
In order for the $k$ transmitted waveforms to coherently add only at the far receiver, each signal is multiplied by the conjugate of the complex exponential (or weights), which provides the necessary phase rotation for directing the beam from the distributed array.
Therefore, the beamforming weights of each transmitter, $k {\in}\{1,2, \ldots, N\}$ for a known receiver at $\left(r, \psi_{rx}, \zeta_{rx} \right)$ is set to $w_k(\psi_{rx},\zeta_{rx}){=}e^{-j\frac{2 \pi}{\lambda} d_k(\psi_{rx}, \zeta_{rx})}$, where $\lambda$ is the wavelength of the carrier. 

The corresponding array factor, given the transmitter locations $\boldsymbol{a}=\left[a_{1}, a_{2}, \ldots, a_{N}\right]$, 
$\boldsymbol{\phi}=\left[\phi_{1}, \phi_{2}, \ldots, \phi_{N}\right] {\in}[-\pi, \pi]^{N}$, and $\boldsymbol{\theta}{=}\left[\theta_{1}, \theta_{2}, \ldots, \theta_{N}\right] {\in}[-\pi, \pi]^{N}$ is defined by \eqref{eq:F}, 

\begin{align} 
F(\psi, \zeta | \boldsymbol{a}, \boldsymbol{\phi}, \boldsymbol{\theta}) &=\frac{1}{N} \sum_{k=1}^{N} \left[w_k(\psi_{rx},\zeta_{rx})\right]  e^{j \frac{2 \pi}{\lambda} d_{k}\left(\psi, \zeta\right)} \nonumber \\
&=\frac{1}{N} \sum_{k=1}^{N} e^{j \frac{2 \pi}{\lambda}\left[d_k(\psi, \zeta)-d_{k}\left(\psi_{rx}, \zeta_{rx}\right)\right]} 
\label{eq:F}
\end{align}

The corresponding far-field beampattern is defined by (\ref{eq:P}),

\begin{equation} 
P(\psi, \zeta |\boldsymbol{a}, \boldsymbol{\phi}, \boldsymbol{\theta}) \triangleq|F(\psi, \zeta |\boldsymbol{a}, \boldsymbol{\phi}, \boldsymbol{\theta})|^{2}
\label{eq:P}
\end{equation}


%
%
%
Consequently, the beampattern is maximum when $(\psi, \zeta)=(\psi_{rx}, \zeta_{rx})$, i.e. along the true direction of the receiver. This is because the phase components of the weights and the impending signal at the receiver in \eqref{eq:F}, are exactly opposite in phase, and therefore add up constructively when $(\psi, \zeta){=}(\psi_{rx}, \zeta_{rx})$, and add up destructively when $(\psi, \zeta)$ deviates from $(\psi_{rx}, \zeta_{rx})$.

\vspace{15pt}
\section{Distributed Receiver Beamforming}
\label{sec:RBF}

{\abbrev} uses a novel formulation of distributed receiver beamforming to perform DoA estimation that is distinct from transmitter beamforming (Section \ref{sec:Distributed}) in the following ways:
\useshortskip
\begin{figure}[h]
        \centering
		\includegraphics[width=.7\linewidth]{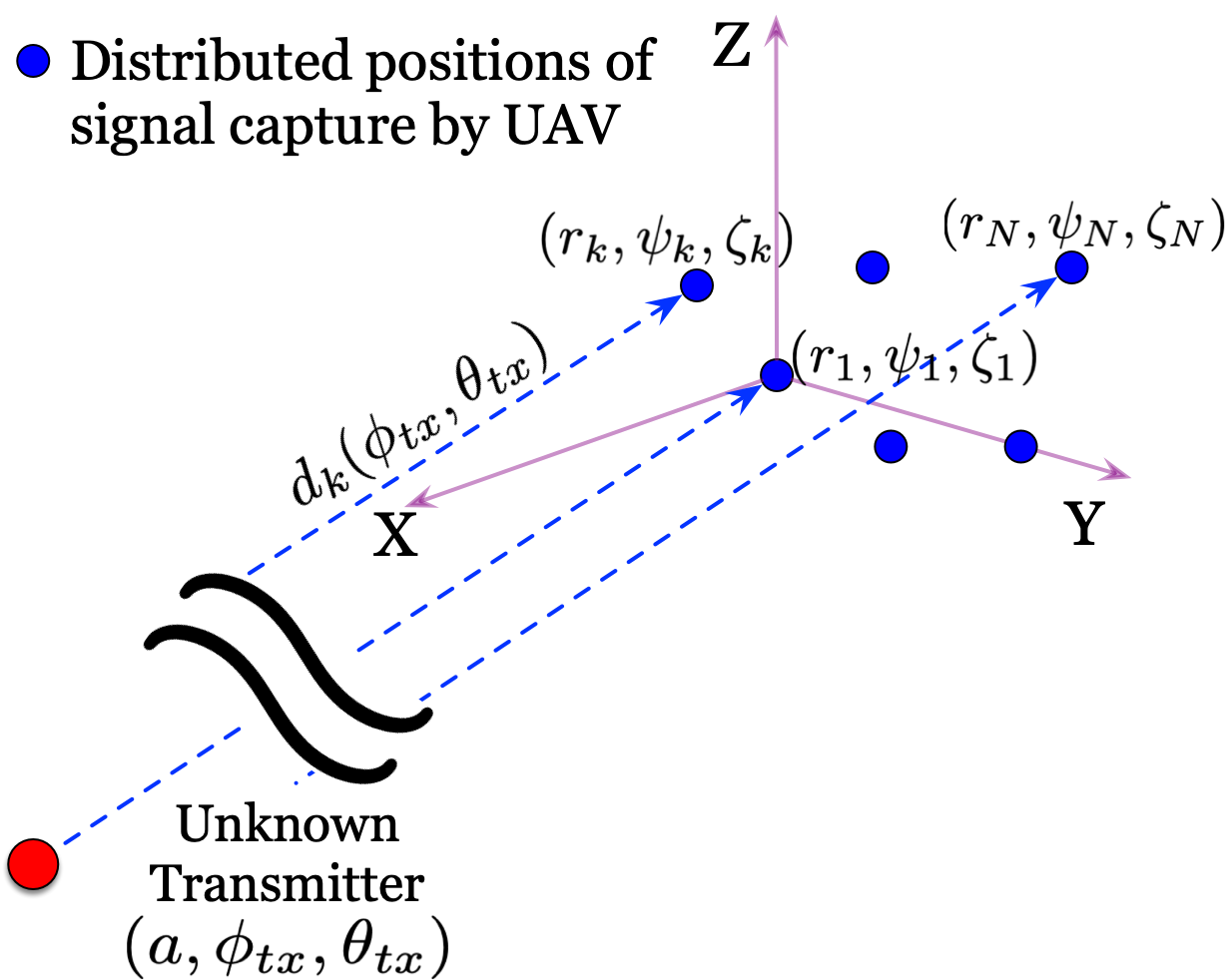}
        \caption{Distributed receiver beamforming}  
        \label{fig:receive_geometry}
\end{figure}
\begin{itemize}
    \item Distributed transmitter beamforming forms a phased array when multiple nodes 
    transmit synchronously to a known receiver. 
    In contrast, {\abbrev} creates a virtual phased array at the receiver by capturing signals from a remote, unknown transmitter 
    at multiple distributed positions to estimate the DoA and its location. 
    \item Transmitter beamforming relies on the knowledge of two parameters: a) The exact location of the far-field receiver $(r, \psi_{rx}, \zeta_{rx})$ and b) Precise locations of the transmitter nodes $(a_{k}, \phi_{k}, \theta_{k})$ to determine the beamforming weights.  
    In {\abbrev}, these variables are unknown and require additional signal processing for receiver beamforming to be practical.
    \item Transmitter beamforming requires precise timing and carrier frequency synchronization for collaborative transmissions. In {\abbrev} the signals are captured asynchronously by a single receiver and aligned in time to create a synchronous, virtual array.
\end{itemize}

To overcome these challenges, we formulate a beamforming algorithm with asynchronous reception from an unknown transmitter using measurable but imprecise positions of spatially distributed array elements. 
Figure \ref{fig:receive_geometry} shows a far-field transmitter and randomly distributed reception points and without loss of generality, we consider the first position of the UAV as the local origin for DoA estimation. 
Keeping the notation in accordance to Section \ref{sec:Distributed}, the coordinates of the UAV is denoted by $pos_{k}{=}(r_k, \psi_{k}, \zeta_{k})$, where $0 {\leq} r_k {<} R$, $- \pi {\leq} \psi_k,\zeta_k {<} \pi$, and $R$ is the radius of the sphere within which the UAV hovers at each location as shown in figure \ref{fig:system}. $pos_k$ is measured using an on-board high precision GPS described in Section \ref{sec:implementation}.
%
Similarly, the transmitter is located at $\left( a,\phi_{tx},\theta_{tx}\right)$, where $a$ is the Euclidean distance between the transmitter and the assumed local origin. Therefore, the Euclidean distance between $pos_k$ and any node at $(a, \phi, \theta)$ is given by \eqref{eq:RxD},
\begin{align}
d_k(\phi, \theta)&\approx a {-} r_{k} \left(\sin \phi\sin \psi_k  \cos \left(\theta{-}\zeta_k\right){+}\cos \phi \cos \psi_k\right) \nonumber\\
&\approx a {-} d'_k(\phi, \theta)
\label{eq:RxD}
\end{align}
where, $d'_k(\phi, \theta)$ is variable and depends on $pos_k$, and the angles $(\phi,\theta)$. 
In order to determine the receiver beamforming weights, $d'_k(\phi, \theta)$ is used, 
since $a$ is unknown and also because it does not impact the DoA and the eventual fix of the transmitter. We discuss this in Section \ref{sec:DoA}.

\noindent
\textbf{The phase of the received signal: }
There are two factors that affect the phase of the signal received at each $pos_k$: a) The phase-shift due to different propagation distances, $d_k(\phi_{tx}, \theta_{tx})$ 
given by the complex exponential, $e^{j\frac{2 \pi}{\lambda} d_k(\phi_{tx}, \theta_{tx})}$, and b) 
Common phase noise induced by carrier frequency offset, frequency selective fading and other native receiver non-linearities. Since the UAV captures distinct packets at different positions, it is necessary to preserve only the phase changes due to propagation delays between the positions (figure \ref{fig:receive_geometry}). This is achieved by identifying and extracting a unique signature embedded in the signal to align the signals captured at different reception points as explained in Sections \ref{sec:feature} and \ref{sec:signal_acq_cond}. The time aligned signals collectively form a synchronous, virtual array where the phase of the signal at $pos_k$ is given by $\frac{2 \pi}{\lambda} d_k(\phi_{tx}, \theta_{tx})$.
Therefore the array factor, given the reception points $\boldsymbol{r}=\left[r_{1}, r_{2}, \ldots, r_{N}\right] {\in} [0, R]^{N}$, $\boldsymbol{\psi}=\left[\psi_{1}, \psi_{2}, \ldots, \psi_{N}\right] {\in}[-\pi, \pi]^{N}$, and $\boldsymbol{\zeta}=\left[\zeta_{1}, \zeta_{2}, \ldots, \zeta_{N}\right] {\in}[-\pi, \pi]^{N}$, is given by \eqref{eq:RxF}, 
\begin{align} 
F(\phi, \theta | \boldsymbol{r}, \boldsymbol{\psi}, \boldsymbol{\zeta}) 
=\frac{1}{N} \sum_{k=1}^{N} \left[w_k(\phi,\theta)\right] e^{j\frac{2 \pi}{\lambda} d_k(\phi_{tx}, \theta_{tx})} 
\label{eq:RxF}
\end{align}
The corresponding beampattern is given by \eqref{eq:RxP},
\begin{equation}
P(\phi, \theta | \boldsymbol{r}, \boldsymbol{\psi}, \boldsymbol{\zeta}) \triangleq|F(\phi, \theta | \boldsymbol{r}, \boldsymbol{\psi}, \boldsymbol{\zeta})|^{2}
\label{eq:RxP}
\end{equation}
Furthermore, the complex weights $w_k(\phi,\theta) = e^{j \frac{2 \pi}{\lambda}d'_k(\phi, \theta)}$, when multiplied with the received signal, rotates the phases of the received signals to constructively and coherently overlap in space, maximizing the beampattern in \eqref{eq:RxP} in the direction of the remote transmitter ($\phi_{tx},\theta_{tx}$), which is also the DoA  
of the signal. The complex weights are explained in Section \ref{sec:DoA}.
\section{System Design}
\label{sec:system}

\noindent

Localization using {\abbrev} has \textit{four} distinct stages: a) \textit{Blind Signature Detection:} The UAV flies to an area of interest and captures the first signal to find a repetitive  signature. b) \textit{Signal Acquisition and Conditioning:} UAV hovers to collect signal traces from multiple positions while recording it's own coordinates. c) \textit{Estimation of DoA:} These signals are used to determine the DoA using the method described in Section \ref{sec:RBF} along with a fusion algorithm to maximize its accuracy and finally,  d) \textit{Transmitter Localization:} The DoA from two nearby locations are used to calculate the coordinates of the unknown transmitter. 

\subsection{Blind Signature Detection}
\label{sec:feature}
Random access based packet transmission is common in many modern wireless protocols, like 802.11~\cite{802_11_spec}, LoRa~\cite{lora} and Sigfox~\cite{sigfox}. 
These protocols use repetitive pattern in the preamble of the packet, which can be detected and extracted by blind detection methods.
This makes {\abbrev} applicable to a wide variety of waveforms unlike prior art~\cite{ToneTrack, Spotfi}, which require specific signal structures. Also, {\abbrev} works without a repetitive pattern as long as the signature (preamble) is known to the receiver. In such cases, the blind signature detection is not required. 

\begin{figure}[t]
        \centering
		\includegraphics[width=.52\textwidth]{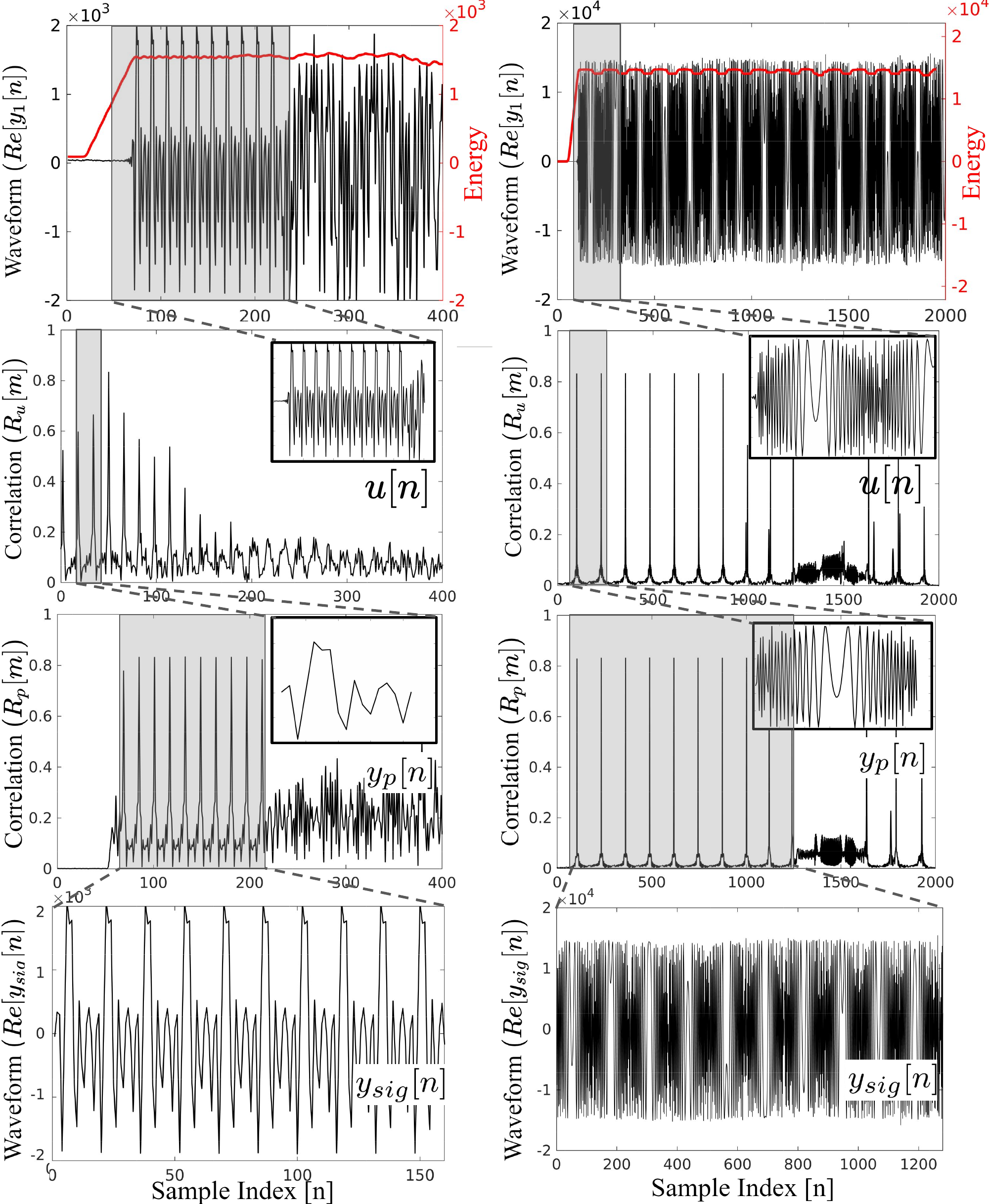}
        \caption{Blind detection and extraction of embedded signature $y_{sig}[n]$ for Wi-Fi (left) and LoRa (right)}  
        \label{fig:detection}
\end{figure}

Let the signal captured at the first position be $y_1[n]$. The goal is to extract a signature, $y_{sig}[n]$ (e.g., short preamble in 802.11 or chirp preamble in LoRa) embedded within $y_1[n]$ that appears in signals captured at all the positions. 
In random access packets, the signature is constructed by repeating a pattern, $y_{p}[n]$ (e.g., one symbol of short preamble in Wi-Fi or a single up-chirp in LoRa). 
%
We present a blind detection scheme, which first extracts the pattern from the signal received at the first position, $pos_1$ of the UAV and then use it to extract the signature. 
This is shown in figure \ref{fig:detection} and works in three steps: \\ 
\noindent
\textbf{Step 1:} The energy envelope of the signal $y_{1}[n]$ is calculated and the first $L$ samples of the envelope above a threshold are extracted, denoted by $u[n]$ in (\ref{eq:energy}),
\begin{align}
\label{eq:energy}
u[n]&=y_{1}\left[\hat{n}{-}L:\hat{n}\right] \quad \text{where,}\\ 
\hat{n}&=\min\left\{n:\left(\frac{1}{L} \sum_{i=0}^{L-1}\Big|y_{1}[n-i]\Big|^{2}\right) \geq \eta_{1}\right\} \nonumber
\end{align}

\noindent where, $\hat{n}$ is the smallest index at which the energy envelope exceeds a decision threshold, $\eta_{1}$ (depends on the lowest target SNR of {\abbrev}).
The value of $L$ ($200$ in this work) is selected so that $u[n]$ contains at least  one instance of $y_{p}[n]$ in an arbitrary waveform. 

\noindent \textbf{Step 2:} $u[n]$ is cross-correlated with the signal $y_{1}[n]$ to determine the exact indexes of 
$y_{p}[n]=y_{1}[m_1{:}m_2]$ using (\ref{eq:top}), 
%
\begin{subequations}\label{eq:top}
\begin{align}
\label{eq:a}
R_u[m]&= \frac{\sum_{i=0}^{L-1} y_1[m+i] u^{*}[i]}{\sum_{i=0}^{L-1}\big|u[i]\big|^2} \\
\label{eq:b}
m_1&=\min\left(\{m:R_u[m] {\geq} \eta_{2} \}\right)  \\ 
m_2&=\min\left(\{m: m{\neq} m_1, R_u[m] {\geq} \eta_{2} \}\right)
\end{align}
\end{subequations}\\
Where, ${m}_1$ and ${m}_2$ denote the two consecutive lowest sample indices at which the normalized cross-correlation $R_u[m]$ exceeds a threshold $\eta_{2}$ (set to 0.5 in this work to minimize false positives). 
The normalized cross-correlation is used for $\eta_{2}$ to be independent of the power level of $y_1[n]$. The correlation also reveals the width of $y_p$ as $W_p{=}{m_2}{-}{m_1}$. 
The cross-correlation 
for Wi-Fi and LoRa is shown in figure \ref{fig:detection} along with the periodic pattern $y_p[n]$.

\noindent \textbf{Step 3:} The pattern $y_p[n]$ is then cross-correlated with the signal $y_{1}[n],$ to extract the signature $y_{sig}[n]{=}y_{1}[m_3{:}m_4+W_p]$ using (\ref{eq:top1}), 
%
%
\begin{subequations}\label{eq:top1}
\begin{align}
\label{eq:a}
R_p[m]= \frac{\sum_{i=0}^{L-1} y_1[m+i] y_{p}^{*}[i]}{\sum_{i=0}^{L-1}\big|y_{p}[i]\big|^2}
\end{align}
\begin{align}
\label{eq:b}
m_3&=\min(\{m{:}R_p[m] \geq \eta_{2} \})  \\ 
m_4&=\max(\{m{:}R_p[m] \geq \eta_{2} \})
\end{align}
\end{subequations}
%
Where, ${m}_3$ and ${m}_4$ denote the lowest and highest sample indices at which the cross-correlation $R_p[m]$ exceeds the threshold $\eta_{2}$. This also reveals the width of the signature, $W_{sig}{=}m_4{+}W_p{-}m_3$ and number of repetitions of $y_p$. In figure \ref{fig:detection}, the cross-correlation of the signal with $y_p[n]$ reveals the complete signature, which contains 10 repetitive patterns for Wi-Fi and LoRa. 
This signature, $y_{sig}[n]$ is used 
to align the asynchronously received signals in time to apply receiver beamforming to compute the DoA. 

\subsection{Signal Acquisition and Conditioning}
\label{sec:signal_acq_cond}

\noindent\textbf{Signal and position capture:}
Signal acquisition is performed by the UAV in several positions within a radius of $1m$ around a location, which are recorded using a high accuracy RTK-GPS~\cite{rtk} positioning system. 
The signal captured at each position is denoted by $y_k[n]$ and the instantaneous position is denoted by $pos_k$.
By not associating to an uniform linear array, {\abbrev} removes the requirement for precise control and movement of the UAV, specially in presence of external factors like wind. 
The output of this process is a set of captured signals along with high-precision GPS locations associated with it for further signal processing.

\begin{figure}
        \centering
		\includegraphics[width=.98\linewidth]{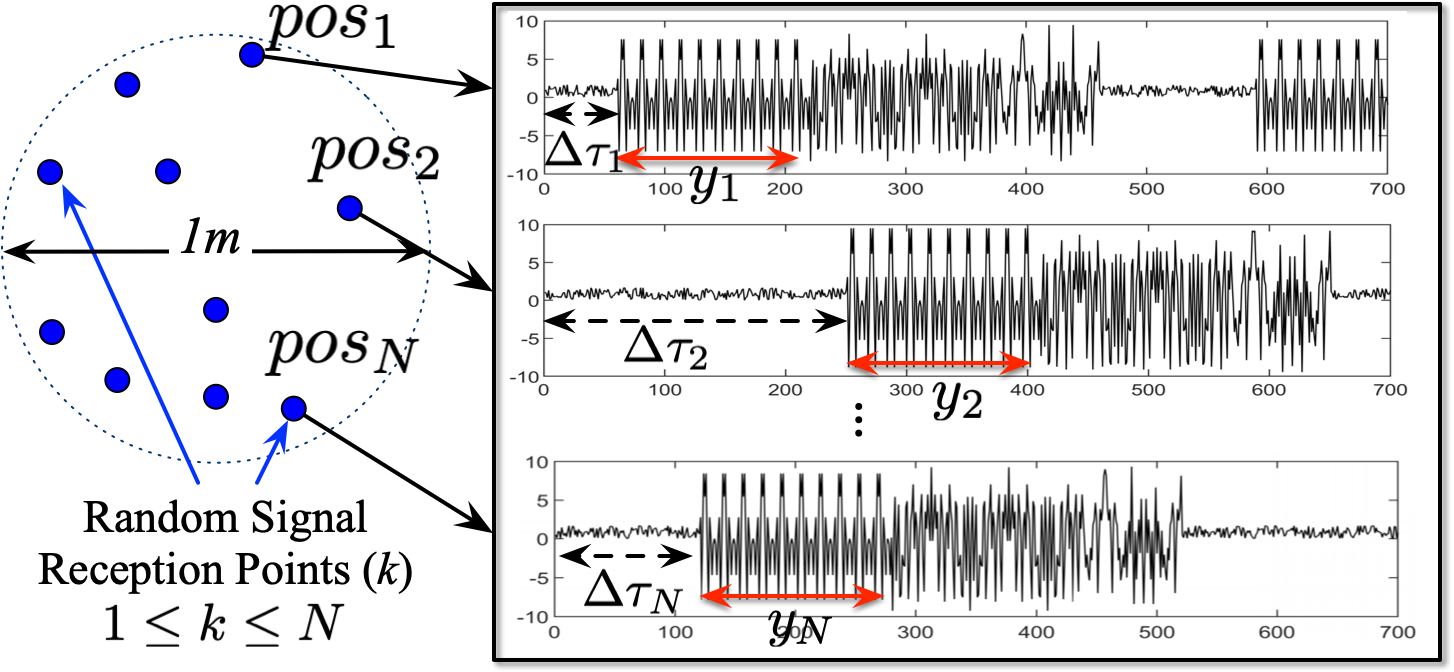}
        \caption{Temporal alignment of $y_k[n]$}  
        \label{fig:align}
\end{figure}

\noindent\textbf{Temporal alignment:}
The UAV captures a continuous stream of the signal transmitted from the remote source at various positions. In each position the signal may contain a packet along with the preamble, but the start of the packet is unknown. 
To utilize the distributed beamforming technique from Section \ref{sec:RBF}, synchronous reception of the same transmitted signal from multiple positions is required. Since we do not use multiple receivers, we emulate this by 
finding the exact indices of $y_{sig}[n]$ within the signals captured from each position and aligning those in time, to generate signals as captured by synchronous receivers. 
Figure~\ref{fig:align} shows the timing alignment 
by cross-correlating the captured signal, $y_k[n]$ with the signature, $y_{sig}[n]$ detected in Section \ref{sec:feature}.
For each position the value of $\tau_k$ for which the cross-correlation in \eqref{eq:tau_k} is maximum, 
is the time-lag of the signal $y_{k}[n]$, 
\begin{equation} 
{\tau}_k=\argmax_{\tau} \left(\sum_{i{=}0}^{W_{sig}{-}1} y_k[\tau+i] y_{sig}^{*}[i]\right) \quad \forall k\in \{1,\ldots N\}
\label{eq:tau_k}
\end{equation}
Now, each signal instance captured at each position can be aligned in time domain 
by shifting the signal instance in time domain by $-\tau_{k}$. 
This is the same as extracting candidate signals, $y_{k}{=}y_k[\tau_k{:}\tau_k{+}W_{sig}]$ that emulate synchronously received signals necessary for receiver beamforming. 
Since, asynchrony corrupts the ToF information of the signal, any approach that relies on ToF information for DoA estimation cannot be employed in this scenario.


\noindent

\noindent\textbf{Carrier frequency offset:}
Since, the same receiver (UAV) is used to capture all the signals, the carrier frequency synchronization is inherently ensured across all signal instances. This obviates the removal of any carrier frequency offset (CFO) between multiple signals, which is critical in multi-receiver distributed beamforming. 
However, CFO may exist between the transmitter and receiver which is estimated and removed using standard techniques~\cite{cfo} at the first reception point. All captured signals are corrected accordingly by the same offset. 

\subsection{DoA Calculation}
\label{sec:DoA}

\begin{figure*}
  \centering
  \subcaptionbox{Random distributed reception points
  \label{fig:array}}[.34\textwidth][c]{%
    \includegraphics[width=.32\textwidth]{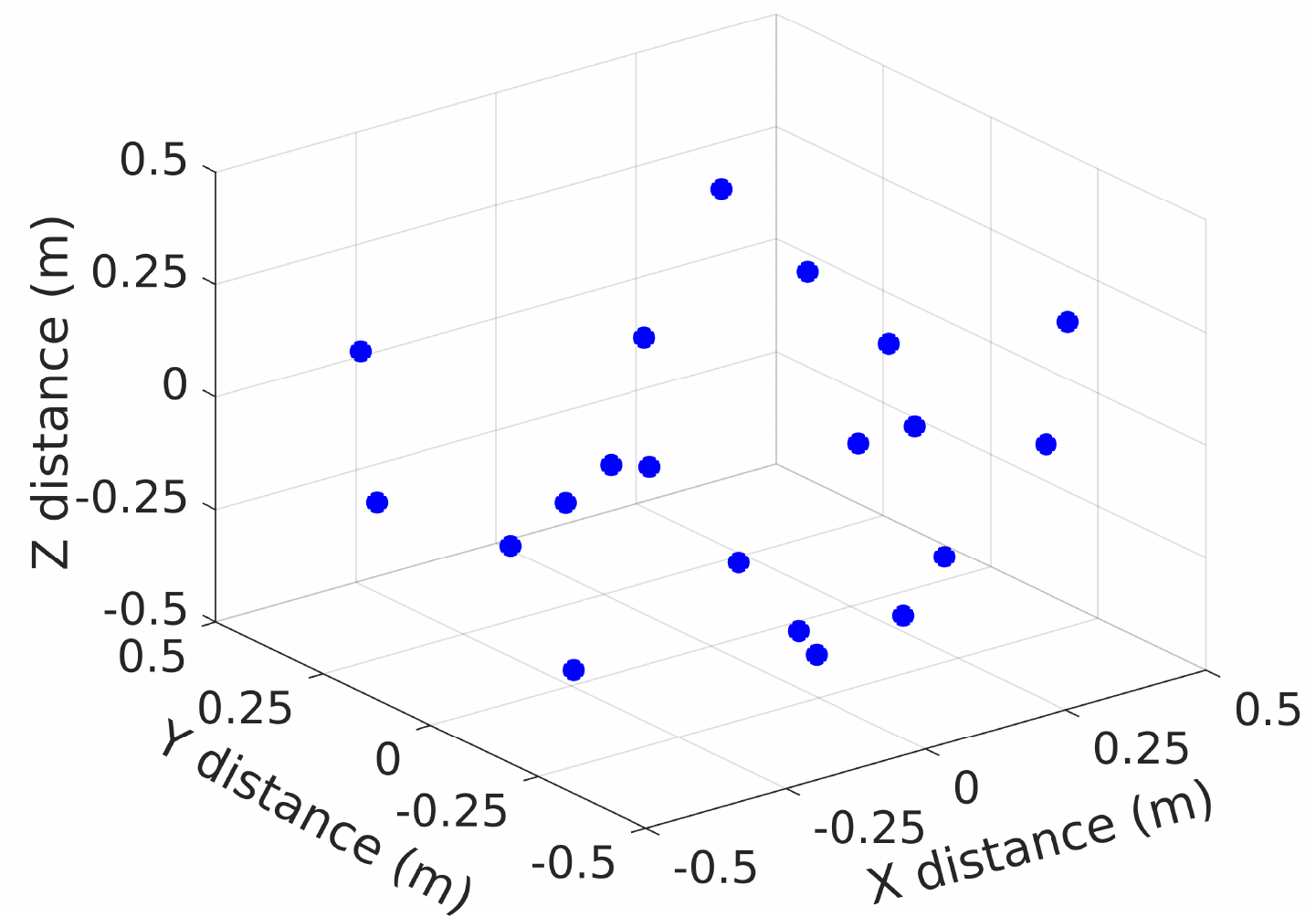}}\enskip
  \subcaptionbox{Beampattern with azimuth angle \label{fig:2dbeam}}[.32\textwidth][c]{%
    \includegraphics[width=.24\textwidth]{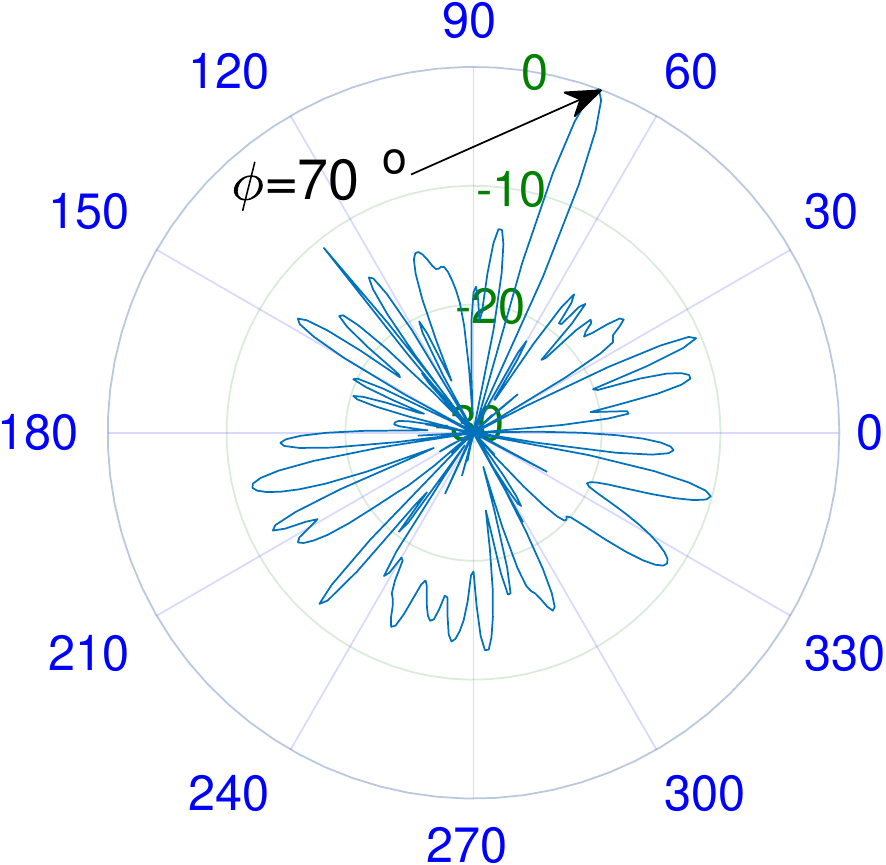}}\enskip
  \subcaptionbox{3D beampattern \label{fig:3dbeam}}[.3\textwidth][c]{%
    \includegraphics[width=.32\textwidth]{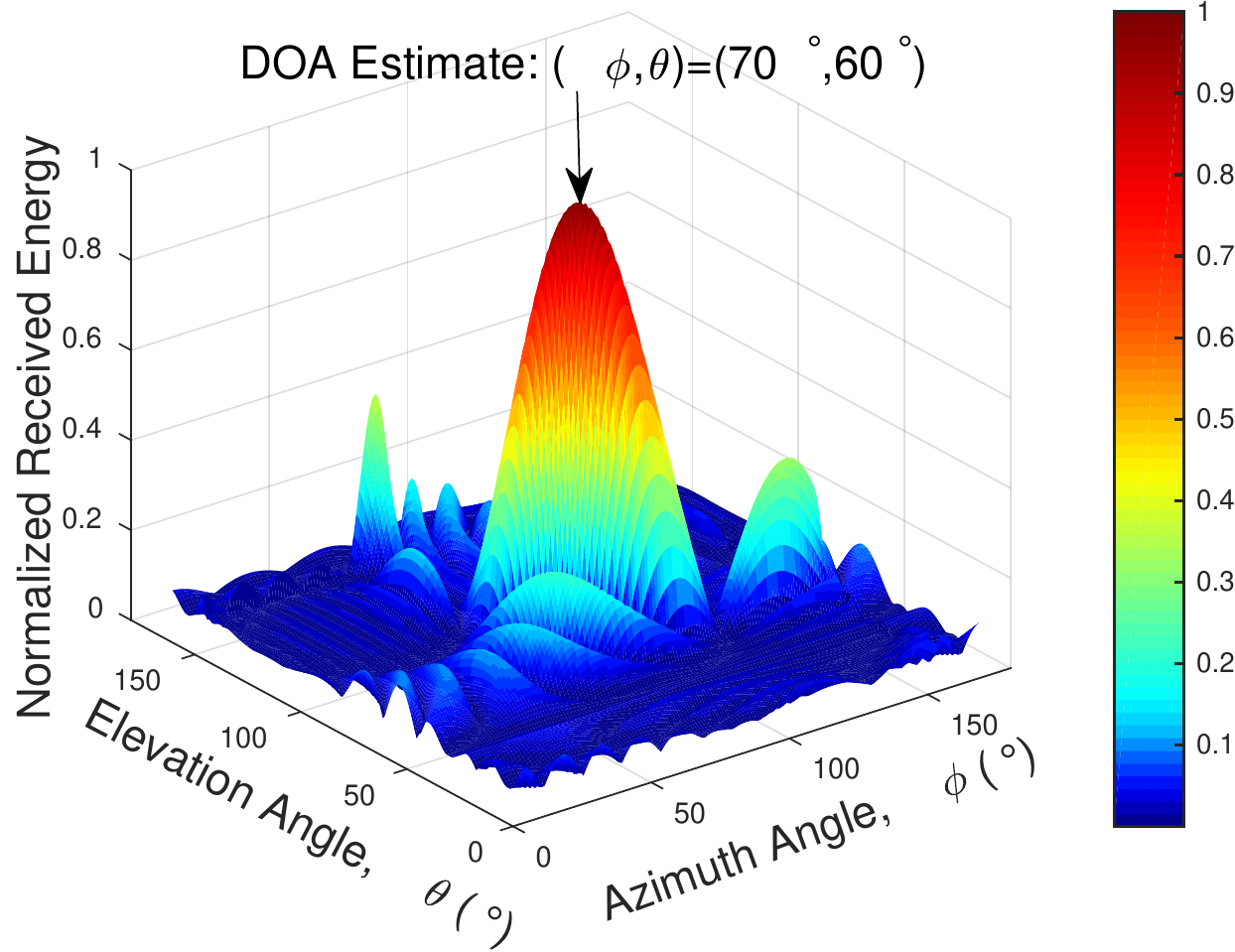}}
    \vspace{5pt}
  \caption{Beampattern for asynchronous distributed receiver array}
  \label{fig:localization}
\end{figure*}

As explained in Section \ref{sec:RBF}, in distributed receiver beamforming
the signal instance at each position is multiplied with an unique complex exponential, 
$w_k(\phi,\theta)$. These weights are a function of the intended direction of the beam ($\phi,\theta$) and the relative position of the reception point $(r_{ k },\psi _{k},\zeta_{k})$. 
Since, the phase of the time aligned signal, $y_k$ is given by $e^{j\frac{2 \pi}{\lambda} d_k(\phi_{tx}, \theta_{tx})}$ 
the weights are chosen to be $w_k(\phi,\theta){=}e^{j\frac{2\pi}{\lambda}d'_{k}(\phi,\theta)}$. 
Multiplying $y_k$ with the complex weight adjusts the phase in a way that the overall beampattern formed by the emulated distributed array is steered in the direction $(\phi,\theta)$.
Now, if this beam is swept in all possible values of $(\phi,\theta)$ where $-\pi{\leq}\phi,\theta{<}\pi$, the received energy is maximum in the direction of the unknown transmitter $(\phi_{tx},\theta_{tx})$.
The array factor is calculated using (\ref{eq:RxF}), 
\begin{align} 
F(\phi, \theta | \boldsymbol{r}, \boldsymbol{\psi}, \boldsymbol{\zeta}) 
&=\frac{1}{N} \sum_{k=1}^{N} e^{j \frac{2 \pi}{\lambda}\left[d'_k(\phi, \theta)+d_{k}\left(\phi_{tx}, \theta_{tx}\right)\right]} \nonumber\\
&=\frac{1}{N} e^{j \frac{2 \pi}{\lambda}a} \sum_{k=1}^{N}  e^{j \frac{2 \pi}{\lambda}\left[d_{k}\left(\phi_{tx}, \theta_{tx}\right)-d_k(\phi, \theta)\right]}
\label{eq:RxFF}
\end{align}
The term $e^{j \frac{2 \pi}{\lambda}a}$ is a constant residual phase in the received signal that does not contribute to the magnitude of the array factor and the corresponding beampattern in \eqref{eq:RxP}. Hence, it has no bearing on the DoA estimation and can be safely ignored.
The signals received at the UAV position, $(\mathbf{r},\boldsymbol { \psi},\boldsymbol { \zeta})$ are used to create a far-field beampattern 
$P(\phi,\theta |\mathbf {r} , \boldsymbol {\psi},\boldsymbol { \zeta})$ and the angles corresponding to its maximum value is the DoA estimate ($\hat{\phi}_{tx},\hat{\theta}_{tx}$) given by \eqref{eq:doa},
\begin{equation} 
(\hat{\phi}_{tx},\hat{\theta}_{tx}) = \argmax_{\phi,\theta} P(\phi,\theta |\mathbf {r} , \boldsymbol {\psi},\boldsymbol { \zeta})
\label{eq:doa}
 \end{equation}
Figure \ref{fig:array} shows an example realization of {\abbrev} with $20$ random positions. 
\note{Figures \ref{fig:2dbeam} and \ref{fig:3dbeam} show the beampattern formed by 
$y_k$ and $pos_k$, where $k{\in}\{1{:}N\}$ 
for a source transmitting from direction $(\phi_{tx},\theta_{tx})=(70^{\circ},60^{\circ})$. 
The beampattern is maximum when the sweep angle matches true direction of arrival of the signal, when there exists a direct path of the signal between the source and the UAV.} 

\subsubsection{Improving the DoA Accuracy}
\label{sec:improve}

\begin{algorithm}[t]
\DontPrintSemicolon
\KwData{$N$, $\mathcal{N}_{max}$, $R_{th}$, set of signals and positions $\{y_{k},pos_{k}\}^j$, for location $j \forall \{1,2\}$, where $k \forall \{1..N\}$}
\KwResult{$(\hat{\phi}_{tx,j},\hat{\theta}_{tx,j})=$ Direction of arrival at location $j$} 
    \For{each location $j \in \{1..2\}$}
    {
        $\mathcal{D}=[\ ]$\;
        \For{$M=(N{-}2){:}{-}1{:}\ceil[\big]{N/2}$} 
        {
            $tmp=\infty$, $exit=0$\;
            \For{$n=1:\mathcal{N}_{max}$}
            {
                Extract $\mathcal{F}_n {=} \{y_k,pos_{k}\}$, a subset with $M$ elements from $\{y_{k},pos_{k}\}^j$ \;
                \For{each $(\phi, \theta) \in \{-\pi..\pi\}$}
                {
                    Calculate $P({\phi, \theta})$ corresponding to $\mathcal{F}_n$ using (\ref{eq:RxP})\;
                    \If{$-3\text{dB}\leq P(\phi,\theta)\leq 0\text{dB}$}{Append $(\phi, \theta)$ to $\mathcal{D}$}
                }    
                $[\bar{\phi}_r, \bar{\theta}_r, R_r]_{r{=} dominant\_cluster}$ {=} k-means-clustering($\mathcal{D}$)\;
                \If{$R_r < tmp $}
                {
                    $tmp=R_r$ \;
                    $(\hat{\phi}_{tx,j},\hat{\theta}_{tx,j})=[\bar{\phi}_r, \bar{\theta}_r]_{r = dominant\_cluster}$     
                }
                \lIf{$tmp<R_{th}$}
                {
                    $exit=1$, break
                }
                
            }
            \lIf{$exit=1$}
                {
                    break    
                }
        }
    }
\caption{ {\abbrev}: Direction of Arrival}
\label{algo:doa}
\end{algorithm}

\note{It is common for the DoA estimate to be influenced by multipath and shadowing effects leading to ambiguous location. The accuracy is improved by employing the following steps: a) Using different subsets of $y_k$ to produce multiple DoA estimates, b) Including all beams that are within 3dB of the maximum beam obtained from (\ref{eq:doa}) and c) Collectively, this creates a much larger search space for the DoA that can be iteratively inferred using a clustering algorithm.}

\noindent
\textbf{a) Expand DoA estimates:}
Since, (\ref{eq:doa}) incorporates the phase from all the signals $y_k$, it is impossible to isolate the effect of errors introduced by one or more of these signals in the final DoA estimate. Instead, choosing $M$ random $y_k$'s out of the available $N$ signals the number of DoA estimates can be vastly increased which leads to more accurate location. 
This produces $\mathcal{N}{=}{N \choose M}$ unique subsets each containing $M$ signals that yield $\mathcal{N}$ DoA using (\ref{eq:doa}).
%
%
%
The value of $M$ 
is a trade-off between the accuracy of the individual DoA and the total number of DoA estimates 
while the value of $N$ is a design choice depending on how fast the DoA estimate is required.
Generally, under perfect time and phase synchronization of the signals and error-free positions of the UAV, the directivity the distributed array increases with $M$~\cite{DistributedBF}, leading to a more accurate DoA estimate. 
However, the value of $\mathcal{N}$ is maximum at $M{=}\ceil[\big]{N/2}$, which limits the range of $M$ to $[\ceil[\big]{N/2},N]$. For each $M$, 
we define $\mathcal{N}$ unique subsets denoted by $\mathcal{F}_n {=} \{y_k,pos_{k}\} \text{ where, } \ n {\in} \{1{:}\mathcal{N}\}$, $k {\in} \{1{:}N\}$, $|\mathcal{F}_n| {=} M$ and estimate the DoA for each set. 
However, to limit on-board computation the number of DoA calculations can be limited to specific number $\mathcal{N}_{max}$, which is determined empirically in Section \ref{sec:parameter_space}.
\begin{figure}
\centering
        \begin{subfigure}[b]{0.25\textwidth}
		\includegraphics[width=\textwidth]{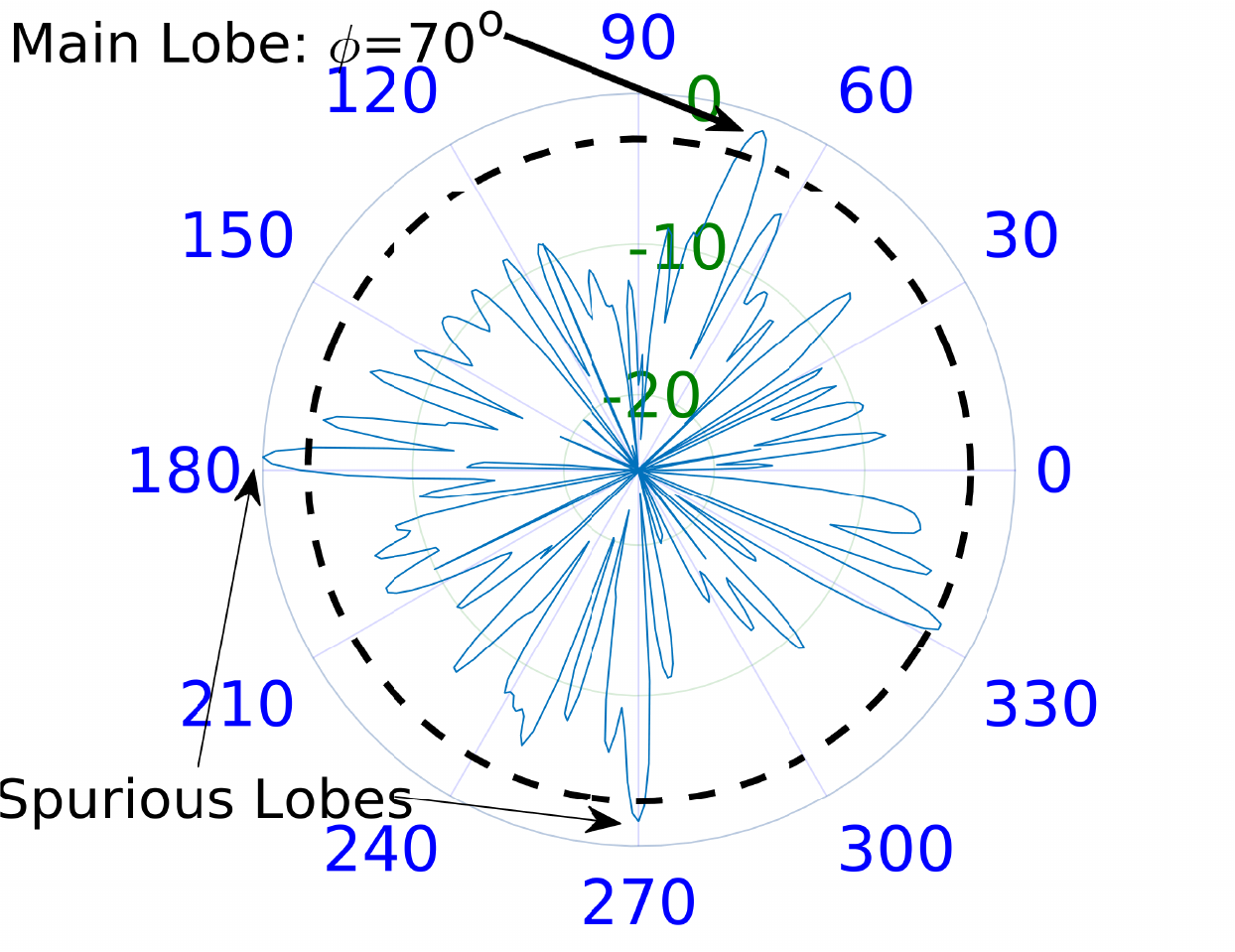}
                \caption{Beampattern with side lobes}
                \label{fig:spurious_beam}
        \end{subfigure}
        \begin{subfigure}[b]{0.23\textwidth}
                \includegraphics[width=\textwidth]{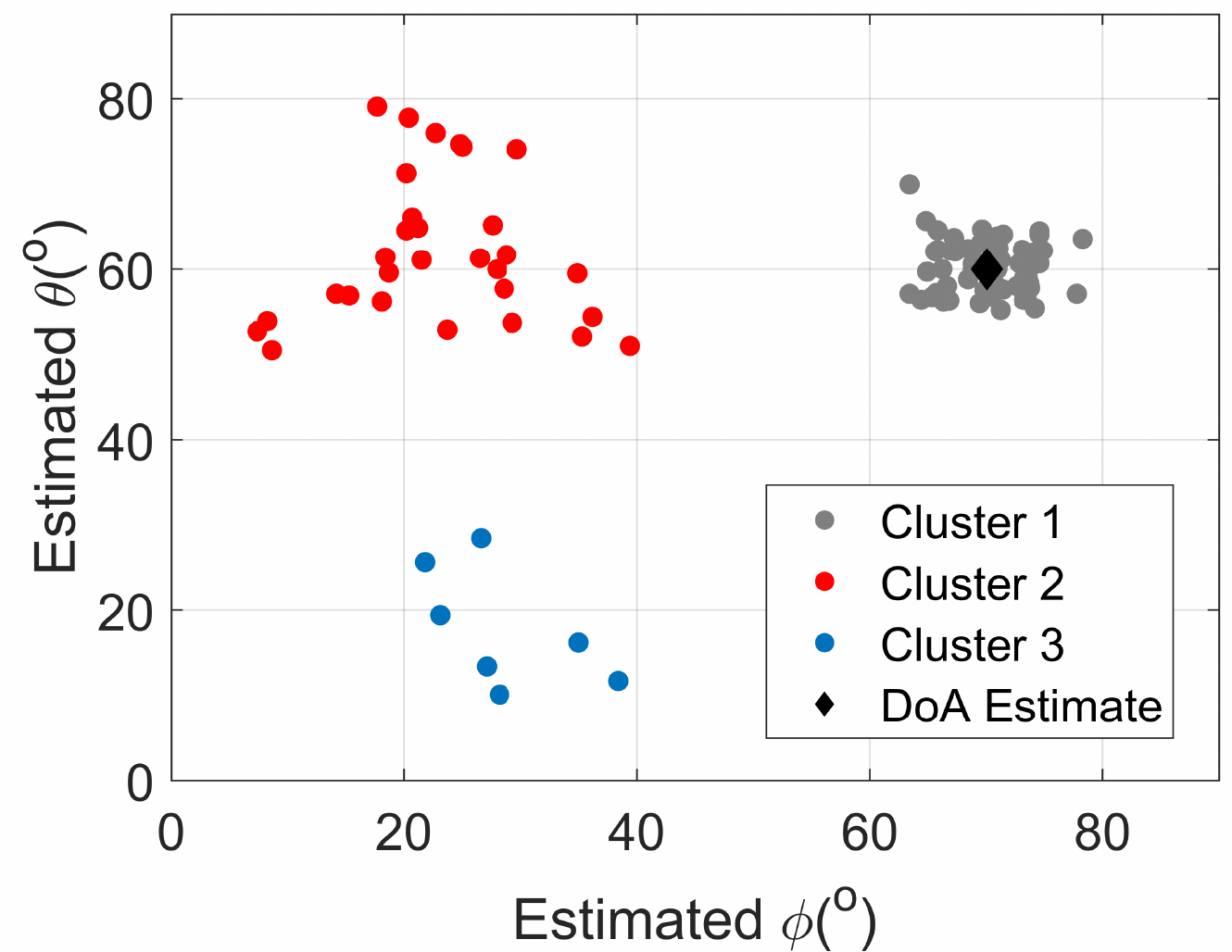}
                \caption{Clustering of DoA}
                \label{fig:cluster}
        \end{subfigure}
\vspace{-10pt}
\caption{(a) Include lobes between 0 to -3dB levels. (b) Median of dominant cluster is the final DoA.}
\vspace{-10pt}
\end{figure}

\noindent
\textbf{b) Consider multiple beam directions:}
In the presence of multipath and shadowing effects in non line-of-sight (NLOS) scenarios, the sidelobes can often lead to ambiguous DoA. For example, figure \ref{fig:spurious_beam} shows spurious lobes other than the main lobe at $(\hat{\phi}_{tx},\hat{\theta}_{tx})=(70^{\circ},60^{\circ})$, which can also be inferred as the true direction. 
To alleviate this error 
we choose all the beam directions that satisfy $-3 \text{dB} \leq P(\phi,\theta) \leq 0 \text{dB (max)}$, 
denoted by the set $\mathcal{D} {=} \{(\hat{\phi}^i_{tx},\hat{\theta}^i_{tx})\}$, where, $i {\in} \{1{:} \text{number of side lobes satisfying the inequality}\}$. 

\noindent\textbf{c) Clustering to improve DoA accuracy:}
\label{sec:cluster}
%
%
From the above discussion, we find that the number of potential DoA can be significantly increased with limited number of signals ($N$). 
However, the DoA may vary for different $\mathcal{F}_n$ because of wireless propagation effects, hardware impairments, timing misalignment or error in UAV coordinates.
Therefore, identifying clusters of $(\hat{\phi}^i_{tx},\hat{\theta}^i_{tx})$ using K-means algorithm reveal regions of high affinity and choosing the median of the dominant cluster provides the most accurate DoA given the $N$ measurements in a location. Algorithm \ref{algo:doa} outlines the steps of clustering, 
represented by the objective function in \eqref{eq:L},
\begin{align} 
\mathcal{L}&{=}\sum_{r{=}1}^{K} \mathcal{L}_r \quad \text{where}, \quad \mathcal{L}_r{=}\sum_{i{=}1}^{|\mathcal{D}|}a_{r}^i\left(\left\|\hat{\phi}_{tx}^i{-}\overline{\phi}_{r}\right\|^{2}{+}\left\|\hat{\theta}_{tx}^i{-}\overline{\theta}_{r}\right\|^{2}\right)
\label{eq:L}
\end{align}
\noindent Where, $K$ is the number of clusters and is kept constant at $3$ and ($\overline{\phi}_{r}$,$\overline{\theta}_{r}$) is the centroid of the $r^{th}$ cluster. 
Here, $a_{r}^i{=}1$ if $(\hat{\phi}^i_{tx},\hat{\theta}^i_{tx})$ belongs to cluster $r$ and $0$ otherwise. 
This objective function is minimized in polynomial time using the algorithm in \cite{kmeans} and the cluster with the smallest radius is considered as the dominant cluster. The centroid of this cluster is the most accurate estimate of the DoA as shown in figure \ref{fig:cluster}. 
At each iteration of Algorithm \ref{algo:doa}, the radius of the dominant cluster 
determined by $R_{r}=\sqrt{\mathcal{L}_{r}/N_{r}}$. 
where $\mathcal{L}_{r}$ is from (\ref{eq:L}) and $N_{k}$ is the number of points in the dominant cluster.
As this radius gets iteratively smaller with the addition of new elements to the set $\mathcal{D}$, it is compared with a threshold, $R_{th}$ (determined empirically in Section \ref{sec:parameter_space}) to avoid redundant iterations that may consume resources on the UAV. Finally, This algorithm is repeated for both locations to get the most accurate DoA possible from the available measurements and is reported in $(\hat{\phi}_{tx,j},\hat{\theta}_{tx,j})$. 

\subsection{Transmitter Localization Using DoA}
\label{sec:localization}

\begin{figure}[h]
\centering
\includegraphics[width=0.8\linewidth]{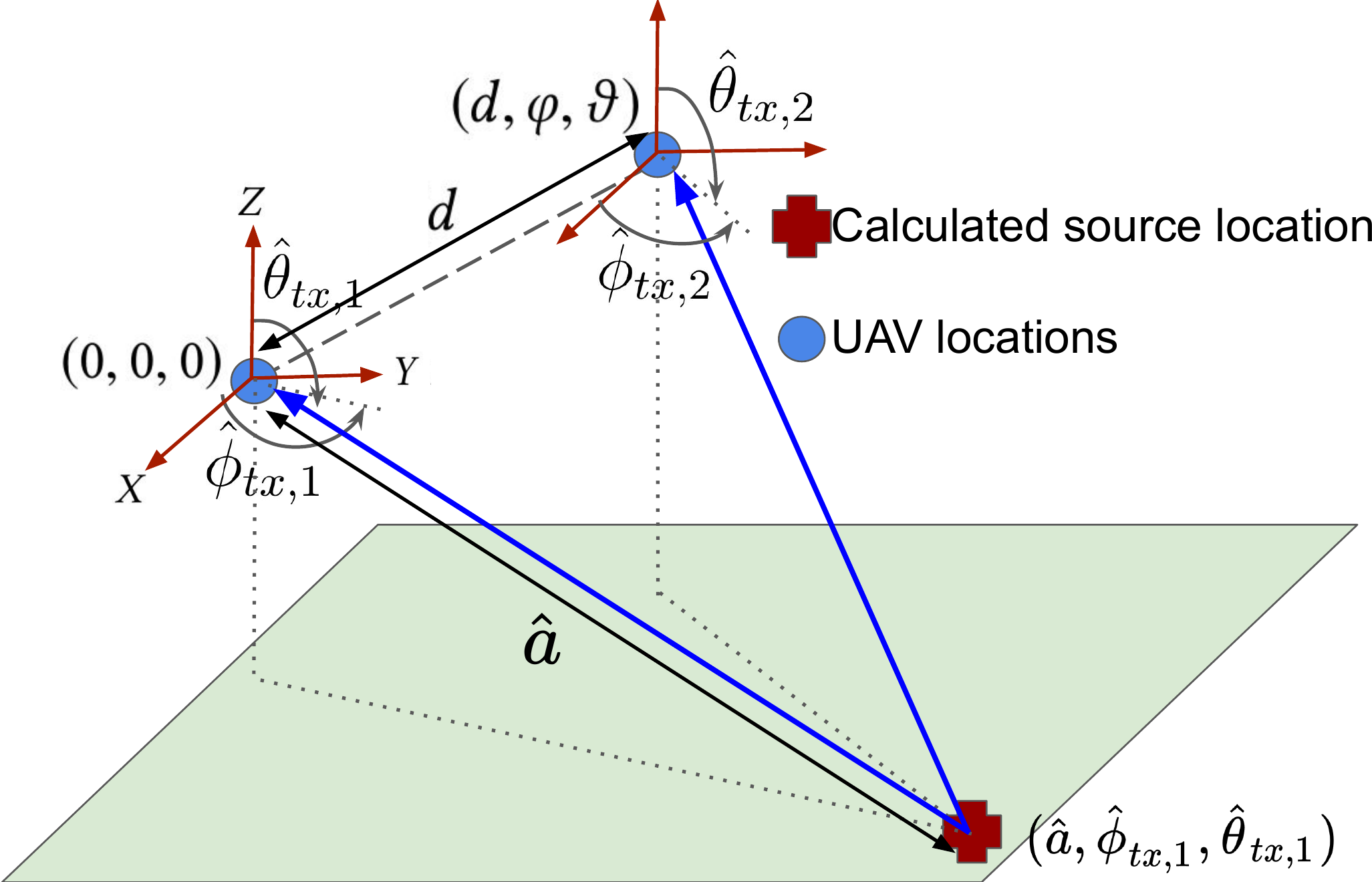} 
\caption{Localization from the UAV}
\label{fig:trigLoc}
\end{figure} 

The unknown transmitter is localized with the DoA obtained from two locations as shown in figure \ref{fig:trigLoc}. The UAV flies to a Location 1 
and calculates the DoA $(\hat{\phi}_{tx,1},\hat{\theta}_{tx,1})$ as described in Section \ref{sec:improve}. Without loss of generality, this location is assumed as the origin with the axes aligned to the GPS axes \cite{loc_book_11}.
Then, it moves to Location 2 
and calculates the DoA $(\hat{\phi}_{tx,2},\hat{\theta}_{tx,2})$. Let the coordinates of Location 2 with respect to Location 1 (the assumed origin) be $(d,\varphi,\vartheta)$.
Since the two locations are known from the RTK-GPS module (Section \ref{sec:implementation}) the coordinates of transmitter is obtained at the intersection of the two lines at angles $(\hat{\phi}_{tx,1},\hat{\theta}_{tx,1})$ and $(\hat{\phi}_{tx,2},\hat{\theta}_{tx,2})$ 
as shown in Figure \ref{fig:trigLoc}. 
The spherical coordinates of the transmitter with respect to the origin, is given by $\textbf{p}_{tx} = (\hat{a},\hat{\phi}_{tx,1},\hat{\theta}_{tx,1})$,  where the range is given by (\ref{eq:range}),
\begin{equation}
    \hat{a}=\frac{d.\sin(\vartheta). \tan(\varphi{-}\hat{\phi}_{tx,2})}{\left(\tan(\hat{\phi}_{tx,1}{-}\varphi)+\tan(\varphi{-}\hat{\phi}_{tx,2})\right)\cos(\hat{\phi}_{tx,1}{-}\varphi)\sin(\hat{\theta}_{tx,1})}
    \label{eq:range}
\end{equation}
Now, if the GPS coordinates of Location 1 is $\textbf{p}_{1}$, then the GPS coordinates of the transmitter is given by $\textbf{p}_{1}+\textbf{p}_{tx}$.
This does not require the transmit power of the source to remain constant unlike commonly required in RSSI based trilateration approaches~\cite{loc_book_11}.

\begin{figure*}[t]
  \centering
  \subcaptionbox{Inaccuracies in temporal alignment of signals ($y_k$) \label{fig:error}}[.24\textwidth][c]{%
    \includegraphics[width=.23\textwidth]{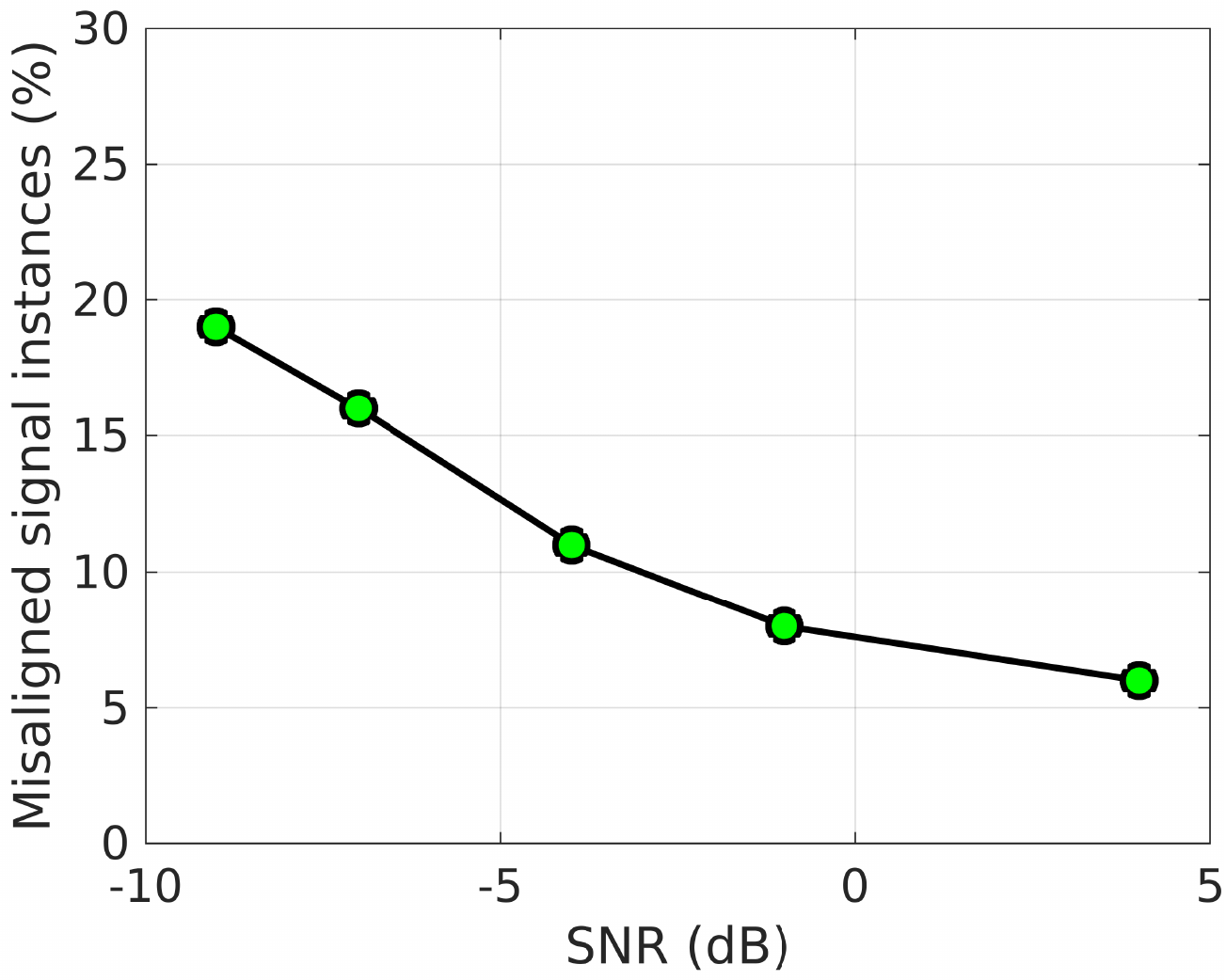}}\enskip
  \subcaptionbox{RTK-GPS: Position error in random trajectory \label{fig:random}}[.24\textwidth][c]{%
    \includegraphics[width=.24\textwidth]{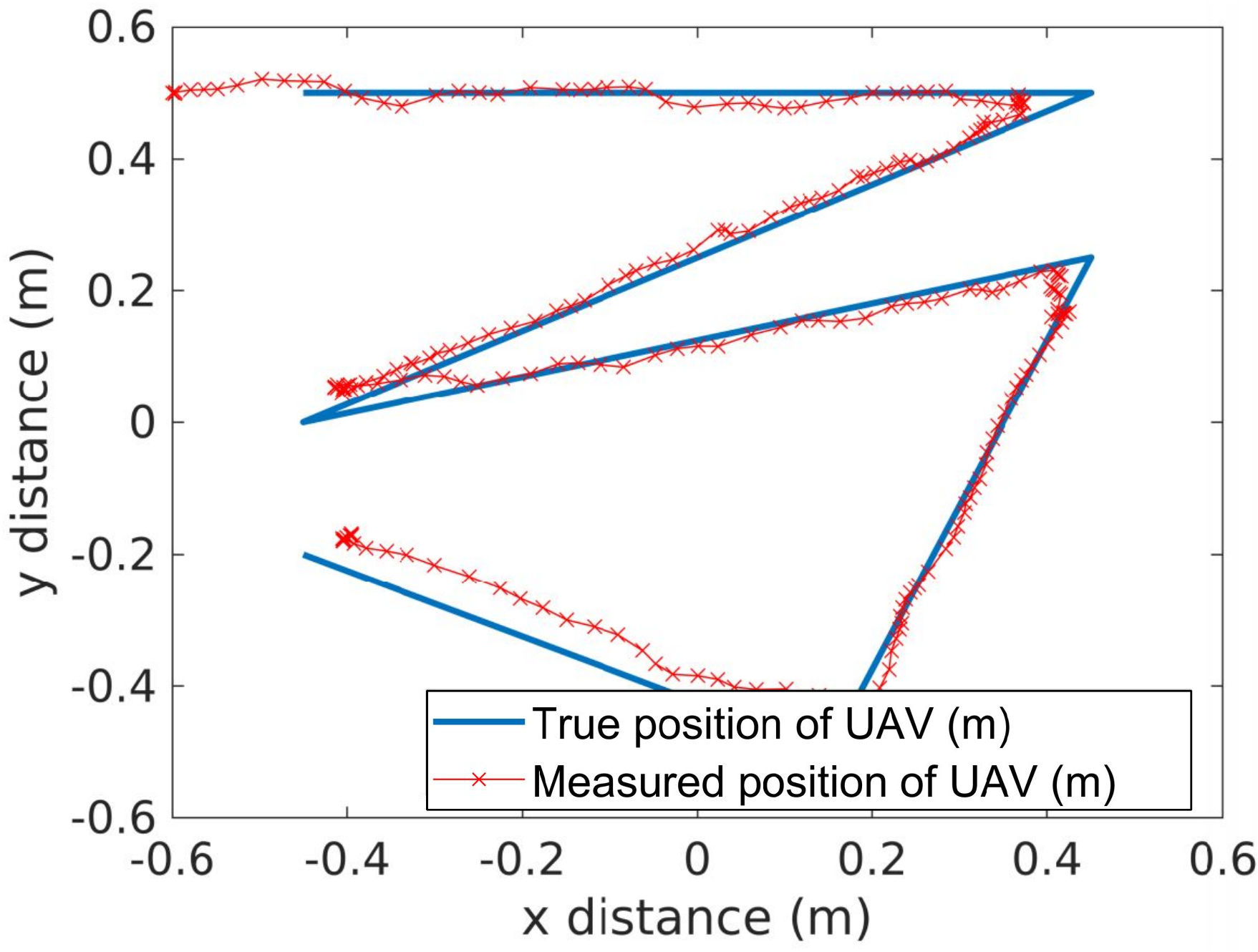}}\enskip
  \subcaptionbox{RTK-GPS: Drift in accuracy over time \label{fig:drift}}[.24\textwidth][c]{%
    \includegraphics[width=.24\textwidth]{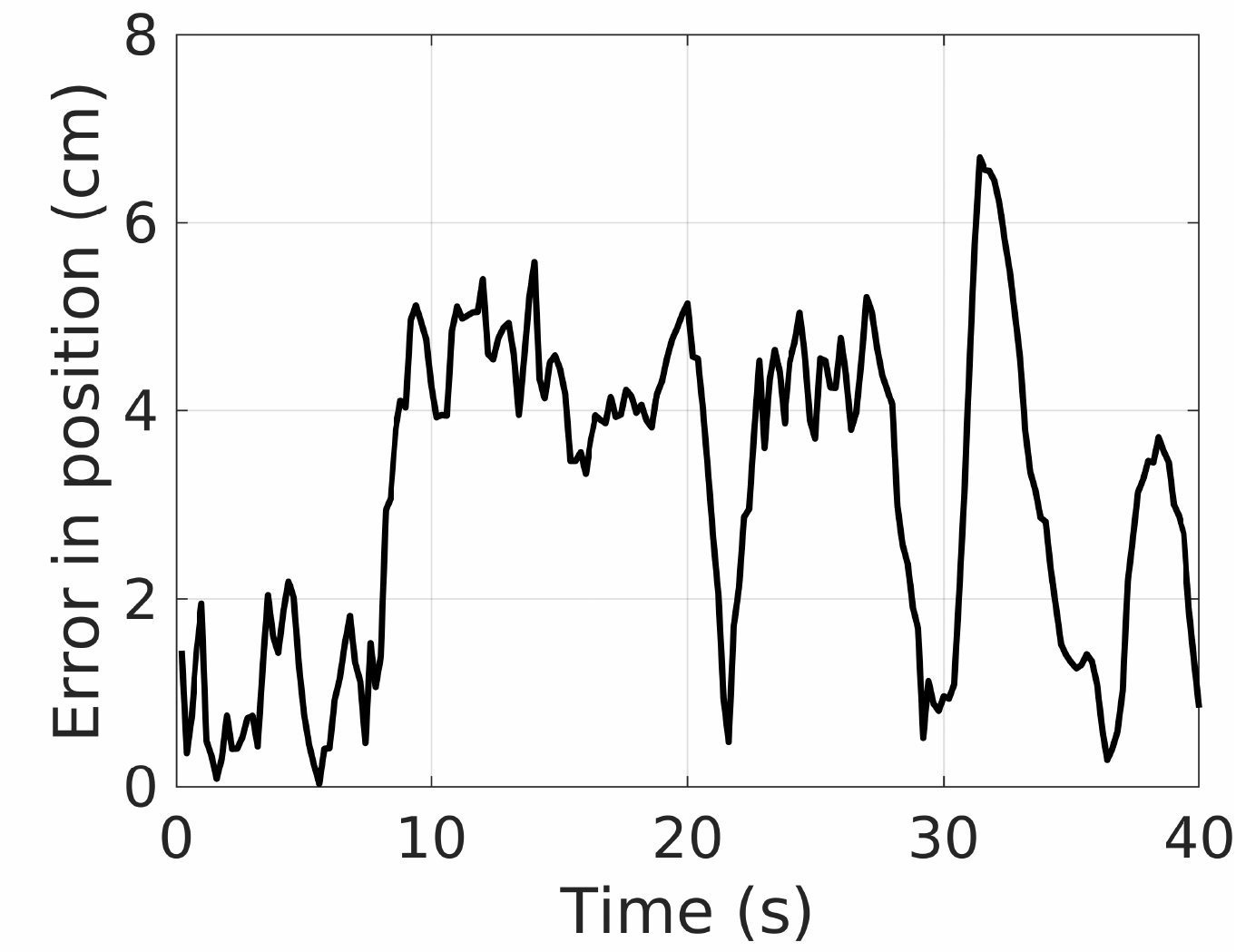}}\enskip
  \subcaptionbox{Error in MUSIC \& {\abbrev} before clustering \label{fig:DoA_MUSIC}}[.24\textwidth][c]{%
    \includegraphics[width=.24\textwidth]{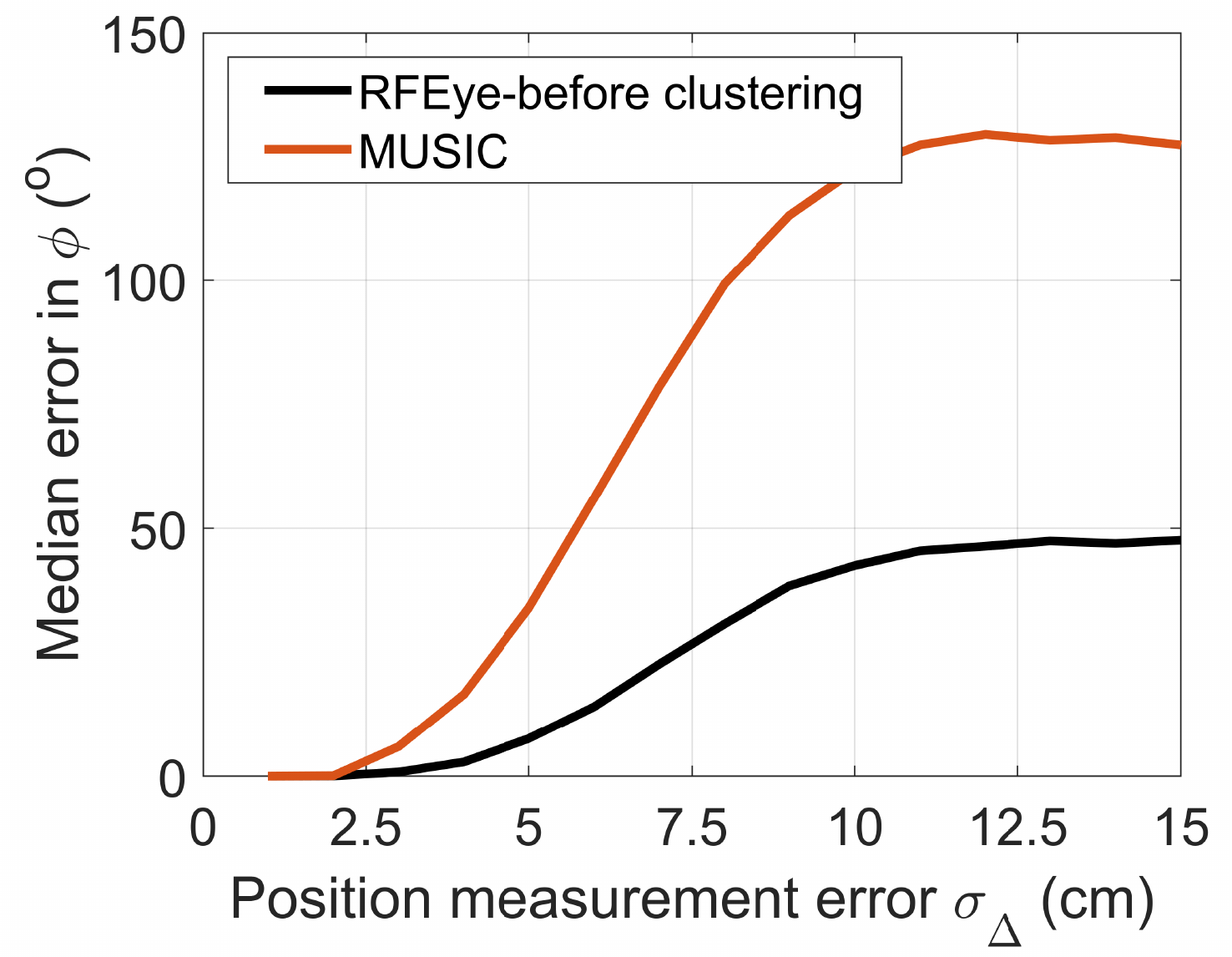}}
  \caption{Error Analysis in {\abbrev} 
  }
  \label{fig:localization}
\end{figure*}
\section{Error Analysis}
\label{sec:error}

The accuracy of {\abbrev} depends on two major sources of errors: a) Imperfect temporal alignment of signal instances $y_k$ and b) Position errors of the UAV. Apart from analyzing the impact of these sources of errors, we also compare the error performance between {\abbrev} and signal subspace based methods~\cite{music_86}.



\noindent \textbf{a) Imperfect temporal alignment:} Imperfect temporal alignment of signal instances manifests as a residual phase error and distorts the beampattern, leading to errors in the DoA. For example in Wi-Fi signals, 1 sample misalignment corresponds to a time shift of $50$ns and $200$ns at sample rates of $20$MHz and $5$MHz respectively. 
Thus the error due to imperfect time alignment is less when the sample rate is high. 
Since the alignment is based on cross-correlation, its accuracy also depends on the SNR of $y_k$ and therefore can cause unwanted correlation peaks using (\ref{eq:tau_k}) at lower values. Figure~\ref{fig:error} shows the misalignment of signal instances with SNR for an outdoor pedestrian channel with Rayleigh multi-path fading profile \cite{rayleigh}. It is evident that the effect of the channel is minimal for SNRs where the packets are actually decodable. Any residual error due to misalignment is eventually removed by the clustering in Algorithm \ref{algo:doa}. 



\noindent \textbf{b) Position errors of the UAV:} The DoA also relies on correct positions of the UAV, measured using the RTK-GPS. Now, values of $pos_k$ can be impaired due to error in RTK-GPS measurement itself, which typically has a cm-level accuracy and precision. Figure~\ref{fig:random} shows the measured positions of the UAV versus the true positions for a random trajectory where the error is greater than the documented accuracy of 2.5cm. Figure~\ref{fig:drift} shows the position error over time. Unlike inertial measurement unit (IMU) positioning systems the drift is not large enough to cause errors in the DoA. Furthermore, the RTK-GPS logs the position 5 times per second. Since, the UAV is in continuous motion during hovering, the exact position may deviate from the last reported value. Since the last reported position is chosen for DoA calculation it can introduce unknown errors. However, {\abbrev} successfully cancels these effects by 
clustering the potential DoA.

\subsection{Error Comparison Between DoA Algorithms}

There are two major classes of techniques employed for 
DoA estimation in RF systems: a) Signal decomposition and b) Array geometry. Algorithms using signal decomposition include the Delay and Sum (DAS) method~\cite{delay_sum_96}, Minimum Variance Distortionless Response (MVDR) beamformer~\cite{mvdr_97}, Estimation of Signal Parameters via Rotation Invariance Techniques (ESPRIT)~\cite{espritAoA89}, MUltiple SIgnal Classification (MUSIC)~\cite{music_86}, Joint Angle and Delay Estimation (JADE)~\cite{jade97}, Space-Alternating Generalized Expectation-maximization (SAGE)~\cite{sage_94}, etc. MUSIC has been widely used in state-of-the-art RF localization systems due to its ability to perform high resolution direction estimation at low SNR. However, it is extremely sensitive to receiver location, gain and phase errors and it requires careful calibration for accuracy. Additionally, the requirement to perform Eigen decomposition and to scan through all the angles is computationally expensive~\cite{MUSIC_Complexity}. This makes the state-of-the-art indoor localization approaches ineffective in resource-constrained agents.  In contrast, {\abbrev} is a light-weight, online and
robust algorithm for DoA estimation. Distributed beamforming~\cite{dist_beam_12} has been theoretically studied in the context of time-synchronized wireless sensors~\cite{dist_loc_10, aoa_14, tDoA_09, toa_14} for uniform~\cite{DistributedBF}, gaussian~\cite{Ahmed} or arbitrary~\cite{Huang} spatial distributions. 
%
We analyze the impact of position error of the UAV on the DoA estimation accuracy based on our formulation of distributed receiver beamforming and compare it to subspace-based techniques like MUSIC.

\subsubsection{{\abbrev} Algorithm}

Let $pos_k$ is represented in rectangular coordinates (assuming coplanar points) by $\left( x _ { k } , y _ { k } \right)$, $k {=} 1, \ldots , N$ is jointly Gaussian with zero mean and variance $\sigma^{2}.$ 
Due to the error in the measured positions, the actual coordinates are $\left( x' _ { k } , y' _ { k } \right), k = 1 , \ldots , N$, where $x' _ { k }{=}x_k{+}\Delta x_k$ and $y' _ { k }{=}y_k{+}\Delta y_k$. 
Here $(\Delta x_k,\Delta y_k)$ is also jointly Gaussian with zero mean and variance $\sigma_{\Delta}^{2}$. Thus, the measured coordinates are jointly Gaussian with zero mean and variance $\sigma'^2{=}\sigma^2{+}\sigma_{\Delta}^{2}$.
In practice, $\sigma_{\Delta}^2 \ll \sigma^2$ because the UAV hovers over a larger radius (1m), 
compared to the magnitude of the error term. Thus, the variance of $pos_k$ is dominated by $\sigma^2$.  
The corresponding polar coordinates, $( r' _ { k }, \psi' _ { k })$, where $r' _ { k } {=} \sqrt { x _ { k } ^ { '2 } {+} y _ { k } ^ { '2 } }$ is Rayleigh distributed and $\psi' _ { k } {=} \tan ^ { - 1 } \frac { y _ { k } } { x _ { k } }$, is uniform in $[-\pi,\pi]$, 

\begin{align*}
f _ { r' _ { k } } ( r' ) {=} \frac { r' } { \sigma^ { '2 } } e ^ { {-} \frac { r ^ {'2 } } { 2 \sigma ^ {'2 } } } , 0 {\leq} r' {<} \infty \quad\text{and}\quad 
f _ { \psi' _ { k } } (\psi') {=} \frac { 1 } { 2 \pi } , {-} \pi {\leq} \psi' {<} \pi.   
\end{align*}
For analysis, assume without loss of generality that $(\phi_{tx},\theta_{tx}){=}(0,\pi/2)$ similar to \cite{Ahmed}. The array factor and the average beampattern over all realizations of $(\boldsymbol{r},\boldsymbol{\psi})$ is given by \eqref{eq:P2}, 
\begin{equation} 
P_{\mathrm{av}}(\phi)=\frac{1}{N}+\left(1-\frac{1}{N}\right)\left|e^{-\frac{\left(4 \pi \sin \left(\frac{\phi}{2}\right)\right)^{2} \sigma^{'2}}{2\lambda^2}}\right|^{2}
\label{eq:P2}
 \end{equation}
The main lobe of the average beampattern is represented by the second term in (\ref{eq:P2}), which is a function of $N$, the steering angle $\phi$ and the distribution of the measured UAV coordinates.
Even in the presence of position errors, the average beampattern is maximum when $\phi{=}0$, i.e., when $\phi$ matches the actual DoA $\phi_{tx}$, showing the robustness of {\abbrev} in the presence of errors. 

\begin{figure*}[t]
	\centering
		\begin{subfigure}[t]{0.35\linewidth}
		 \includegraphics[width=\linewidth]{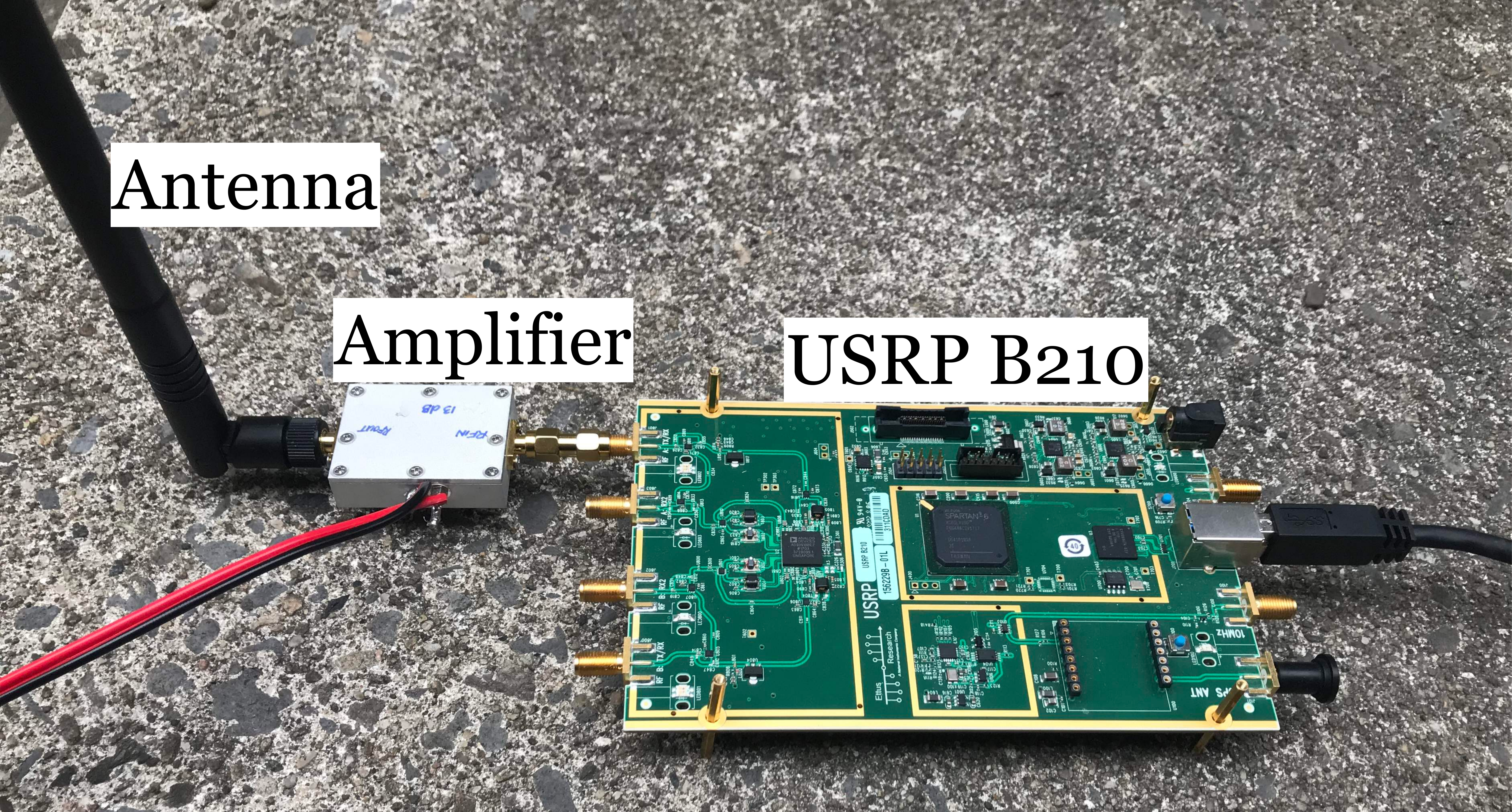}
		\caption{RF source with USRP B210 and 10dBi antenna}
		\label{fig:target}   
	\end{subfigure}
    \enskip 
    \begin{subfigure}[t]{0.245\linewidth}
		\includegraphics[width=\linewidth]{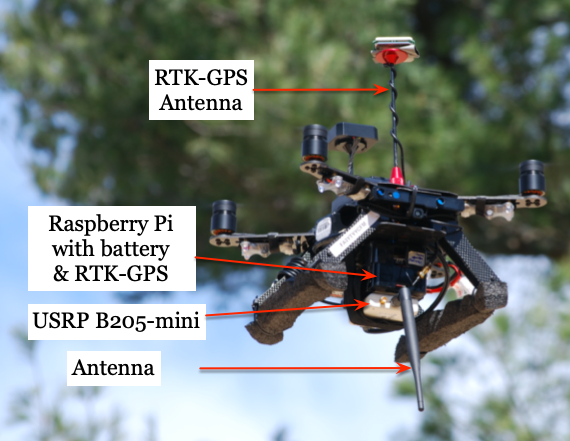}
		\caption{Intel Aero UAV with USRP B205 mini}
		\label{fig:drone}
	\end{subfigure}
    \enskip 
	\begin{subfigure}[t]{0.375\linewidth}
	    \includegraphics[width=\linewidth]{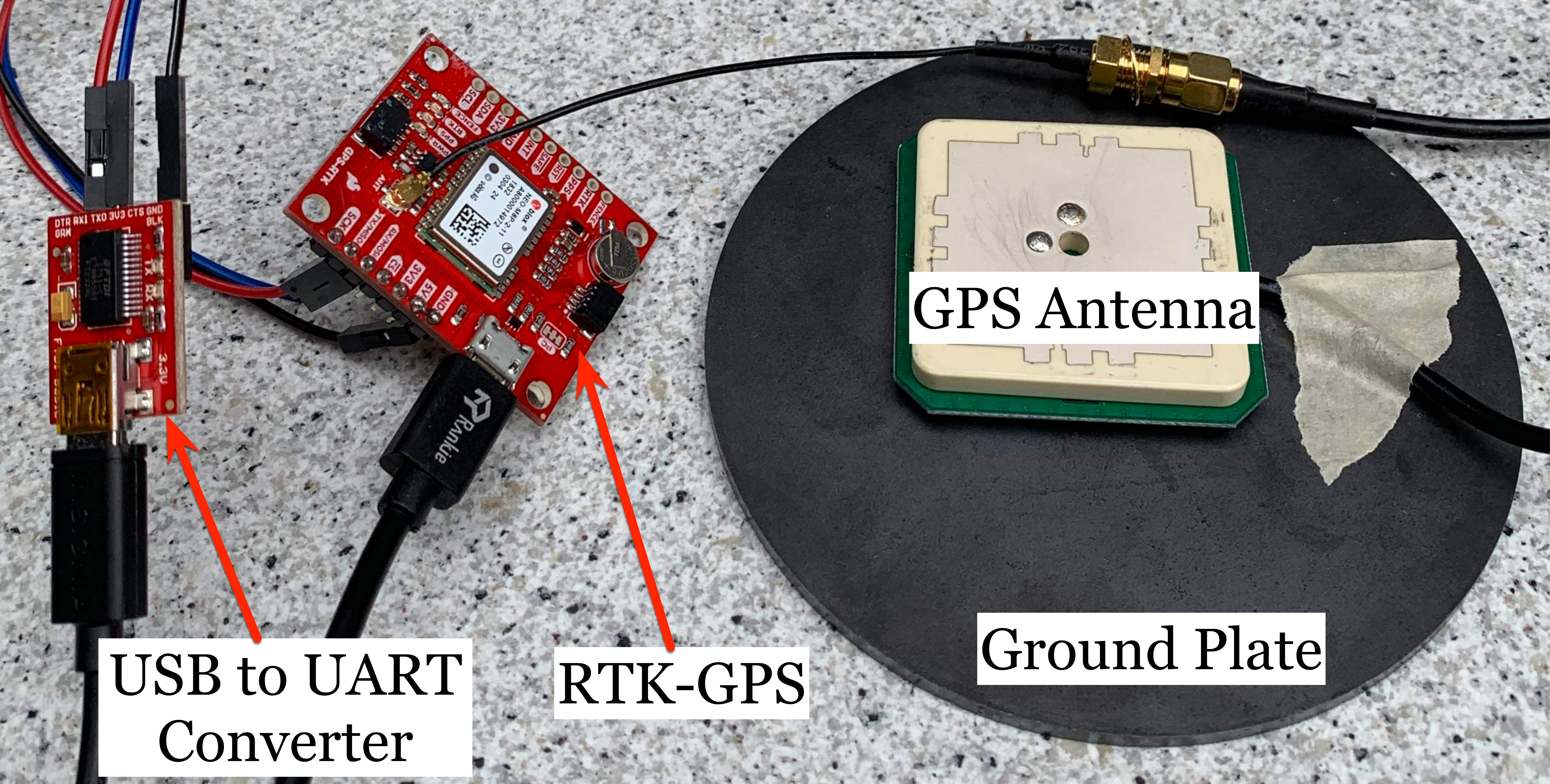}
		\caption{RTK Base with RTK-GPS and GPS antenna on ground plate}
		\label{fig:rtk_base}    
	\end{subfigure}
	\caption{Experimental hardware setup of {\abbrev}} 
	\label{fig:exp_setup}
\vspace{5pt}
\end{figure*}

The main lobe in (\ref{eq:P2}) decays exponentially with a rate proportional to the variance $\sigma^{'2}$. 
Two important observations can be made from this: 1) A narrow distribution of $pos_k$, i.e., small $\sigma^{2}$ yields greater power in the main lobe of the beampattern and more accurate DoA and 2) Since the power of the main lobe is dominated by 
$\sigma^{2}$ 
distributed receiver beamforming is resilient to position errors. However, if $pos_k$ are farther apart (higher $\sigma^{2}$), it will decrease the power in the main lobe  but will have less error in the beampattern. 
Thus, the variance of $pos_k$ is a design trade-off in {\abbrev}.

\subsubsection{MUSIC Algorithm}
\label{sec:music}
The MUSIC spectrum \cite{music_86} measures the distance between the steering vector, $\mathbf{a}(\phi,\theta)$ (the vector of complex weights to steer the beam in direction $(\phi,\theta)$) 
and the noise subspace $\mathbf{E}_{n}$ determined by Eigenanalysis of $y_k$.  
to calculate the DoA using (\ref{eq:music}), 
\begin{align} 
P_{MUSIC}(\phi,\theta)&=\frac{1}{\mathbf{a}(\phi,\theta)^{H} \mathbf{E}_{n} \mathbf{E}_{n}^{H} \mathbf{a}(\phi,\theta)} \label{eq:music}\\
\quad \text{where,}
\quad
\mathbf{a}(\phi,\theta)&=\left[w_1(\phi,\theta),\ldots,w_N(\phi,\theta)\right]\nonumber
 \end{align}
$P_{MUSIC}(\phi,\theta)$ generate peaks when the steering vector is exactly orthogonal to the noise subspace vector, which happens only when ($\phi,\theta$) coincides with the DoA of the incoming signal. 
Hence, the accuracy depends on the orthogonality of $\mathbf{E}_{n}$ and the steering vector $\mathbf{a}(\phi,\theta)$ which is sensitive to the values of $pos_k$ that are prone to measurement fluctuations and errors as shown in figure \ref{fig:localization}a-c. 
This leads to a bias in the DoA estimate from MUSIC proportional to the variance $\sigma_{\Delta}^{2}$ \cite{MUSIC_Asymptotic_Bias}.
\note{Figure \ref{fig:DoA_MUSIC} shows better accuracy of {\abbrev} over MUSIC without clustering even at smaller values of error in $pos_k$.} 

\section{Computational Complexity of {\abbrev}}
\label{sec:complexity}

The computational complexity of {\abbrev} is determined by the four steps in Section \ref{sec:system}.
It is straight-forward to conclude that the blind detection in Section \ref{sec:feature} runs in constant time due the fixed correlation window ($L$), and the temporal alignment in Section \ref{sec:signal_acq_cond} has a linear time complexity of $\mathcal{O}(N)$. 
DoA calculation in Section \ref{sec:DoA} involves extracting $\mathcal{F}_n$ and clustering the potential DoA according to Algorithm \ref{algo:doa}. 
The second loop repeats for a maximum of $(N{-}2){-}\ceil[\big]{N/2}$ times for each location $j$ and the third loop iterates for a maximum of $\mathcal{N}_{max}$ times for each $M$. 
Extracting $\mathcal{F}_n$ 
has a linear complexity of $\mathcal{O}(N)$.
Let, $G{=}360^o{\times}360^o$ denote the search space for the azimuth ($\phi$) and elevation ($\theta$).
Computing the array factor and beampattern using (\ref{eq:RxF}) and (\ref{eq:RxP}) for each $G$ and $M$ reception points ($|\mathcal{F}_n|$) has complexity $\mathcal{O}(MG)$ and determining the direction of maximum power using (\ref{eq:doa}) with $\mathcal{O}(G)$. 
Therefore, {\abbrev} computes the DoA in linear time of $\mathcal{O}((MG{+}G){\approx}\mathcal{O}(MG)$.
In contrast, MUSIC has a quadratic complexity in $M$, $\mathcal{O}(M^2G{+}M^2){\approx}\mathcal{O}(M^2G)$ \cite{MUSIC_Complexity}.
Since, DoA calculation in {\abbrev} has a lower complexity than MUSIC it can execute in real-time on resource-constrained devices like UAV.
The clustering uses a low complexity k-means algorithm that executes in  $\mathcal{O}(K\mathcal{N}_{max}){\approx}\mathcal{O}(\mathcal{N}_{max})$ time \cite{kmeans_complexity}, where $K{=}3$. 
Therefore, the overall complexity of {\abbrev} is given by (\ref{eq:comp}),
\begin{align}
    &\mathcal{O}(\{N{-}2{-}\ceil[\big]{N/2}\}\mathcal{N}_{max}\{N{+}MG{+}\mathcal{N}_{max}\})\nonumber\\
    &\approx\mathcal{O}(N\mathcal{N}_{max}(N{+}MG{+}\mathcal{N}_{max}))
    \label{eq:comp}
\end{align}
%
Since, $\mathcal{N}_{max}$ is constant and $M{<}N$, the overall complexity is $\mathcal{O}(N(N{+}NG)){\approx}\mathcal{O}(N^2G)$. For resource constrained devices, $M$ can be fixed empirically (explained in Section \ref{sec:parameter_space}) and in practice $N{\ll}G$, which leads to a complexity of $\mathcal{O}(NG)$ from \eqref{eq:comp}. Further, the complexity of scanning the search space $G$, can be improved by first scanning at a low resolution and iteratively scanning at higher resolutions only at angles at which the beampattern is maximum, yielding a logarithmic search complexity of $\mathcal{O}(log_2 G)$ \cite{Ubicarse}. Thus, the complexity of {\abbrev} is $\mathcal{O}(Nlog_2 G)$.
\section{Implementation \& Experimental Setup}
\label{sec:implementation}
\begin{figure}
\centering
\includegraphics[width=\linewidth]{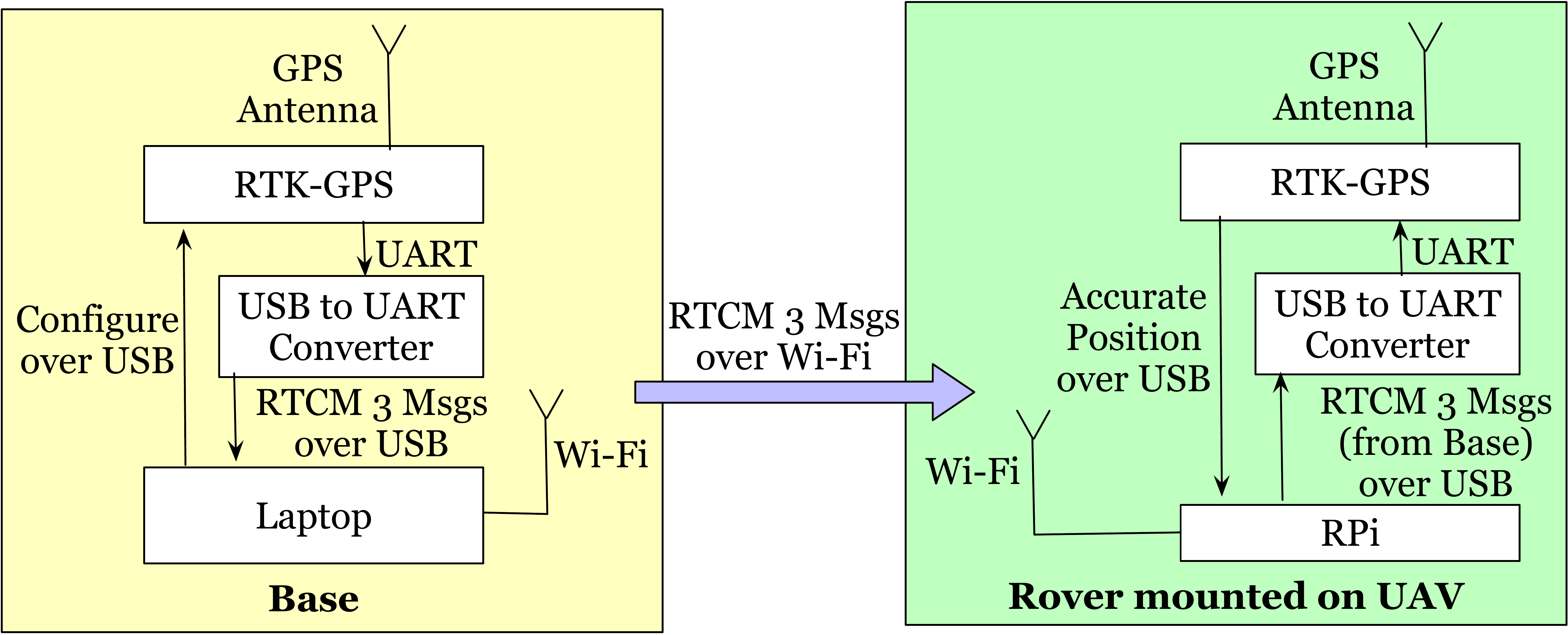}
\caption{RTK Setup Block Diagram}
\label{fig:rtk_block}
\vspace{-15pt}
\end{figure}
\begin{figure*}
  \centering
    \subcaptionbox{Number of reception points \label{fig:NumPoints}}[.3\textwidth][c]{%
    \includegraphics[width=0.9\linewidth]{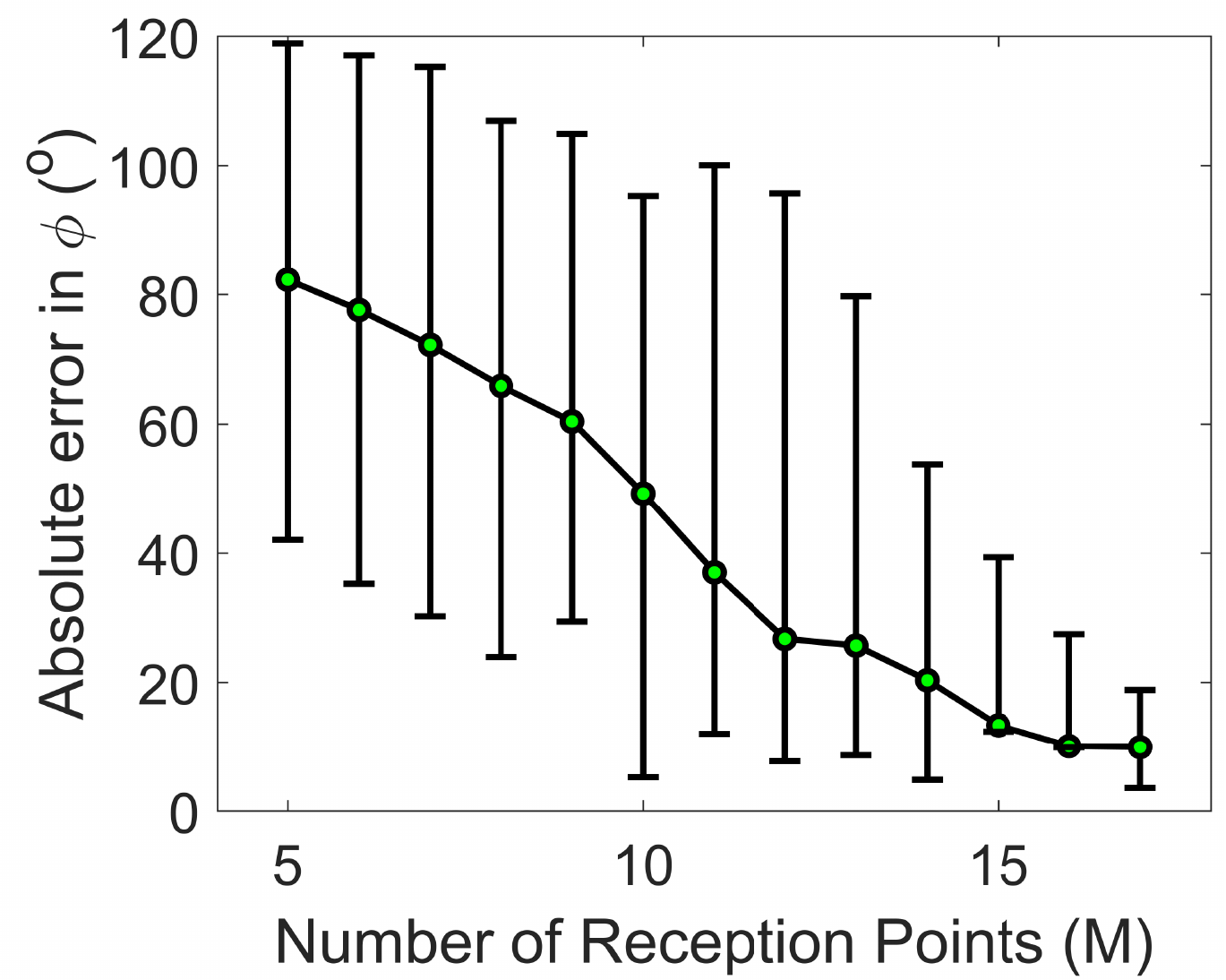}}\quad
    \subcaptionbox{Radius of dominant cluster \label{fig:radius}}[.3\textwidth][c]{%
    \includegraphics[width=0.9\linewidth]{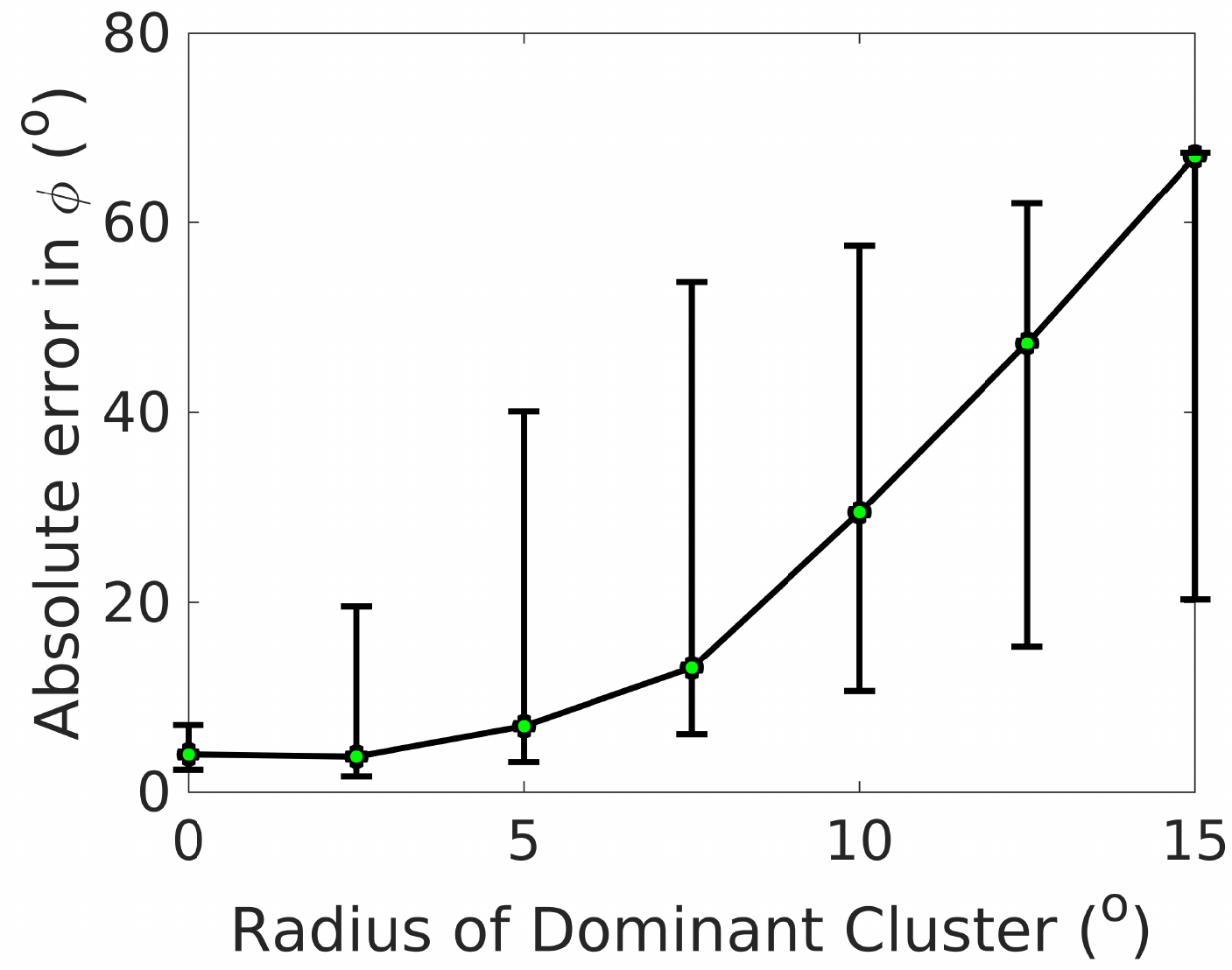}}
    \quad
  \subcaptionbox{Number of $\mathcal{F}_n$ instances \label{fig:NumClusters}}[.3\textwidth][c]{%
    \includegraphics[width=0.9\linewidth]{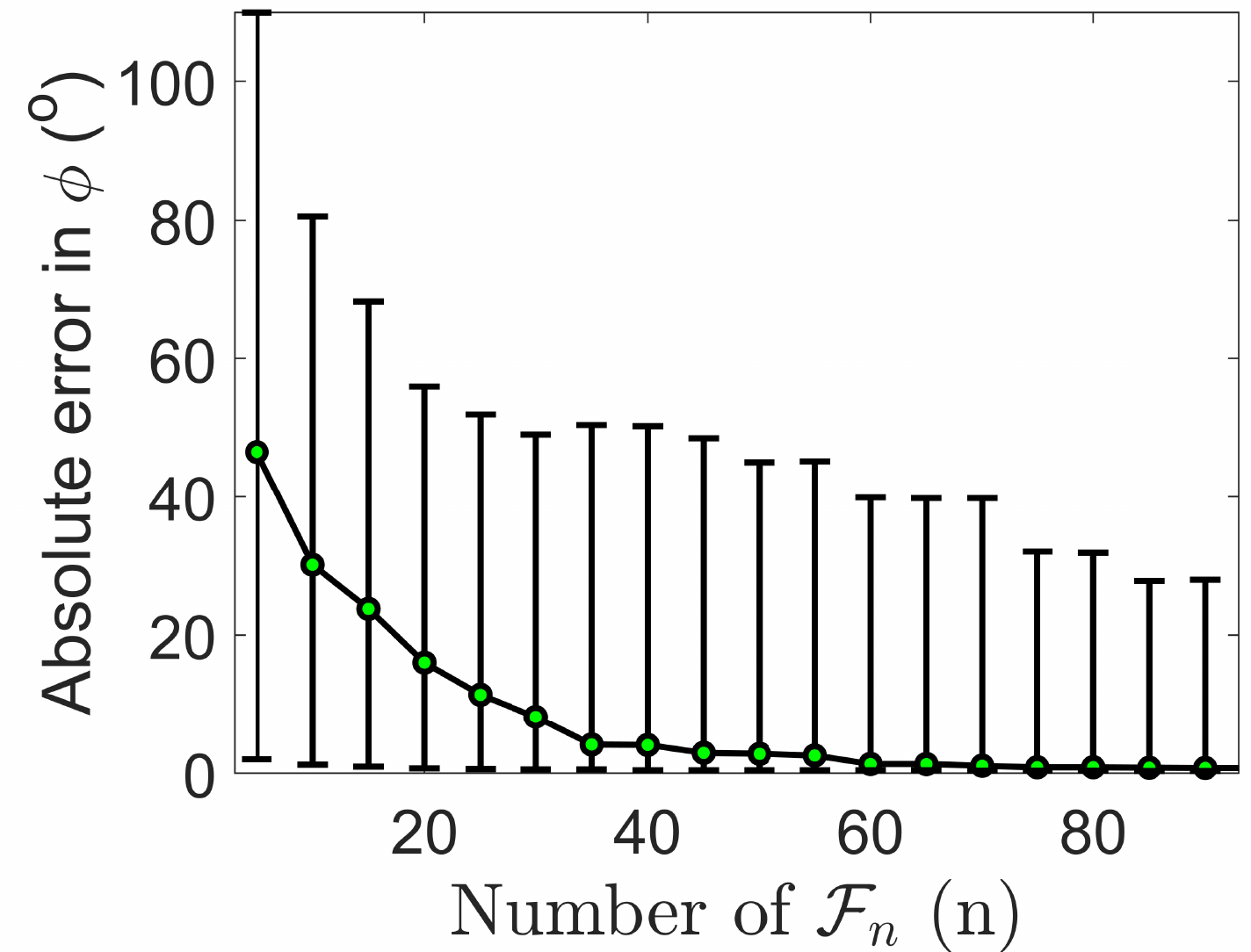}}
  \caption{The impact of tunable parameters on DOA accuracy at SNR=11dB and BW = 20MHz}
  \label{fig:points_pkts_clusters}
\end{figure*}

\label{subsec:signal_recv}
\subsection{Transmitter and UAV hardware}
The transmitter consists of a USRP B210~\cite{B210} with a 10dB RF amplifier and 6dBi antenna to transmit a signal with a repetitive pattern. For experimental purposes the Wi-Fi preamble was continously transmitted which served as the embedded signature, $y_{sig}[n]$. For over-the-air experiments an unused Wi-Fi channel was used. The amplifier boosts the B210 signal to 26dBm which was sufficient to conduct outdoor experiments 
but much lower than commercial Wi-Fi access points and the FCC's recommendation of 
4W in 2.4GHz unlicensed spectrum. 
Figure~\ref{fig:target} shows our hardware setup of the transmitter. 

For the UAV, we use one Intel Aero~\cite{Aero} as the main component of {\abbrev} running Ubuntu. Complex digital samples, $y_k[n]$ are acquired using a USRP B205 mini~\cite{B205} with a single antenna mounted on the chassis as shown in figure~\ref{fig:drone}. The signal is sampled at 5, 10 and 20MHz and streamed to the RAM disk \cite{RAMdisk} memory over USB 3.0. 
The UAV hovers within a sphere of 1m radius and we conduct flight experiments in both windy and calm days to show the performance in all possible scenarios. The USRP B205 mini is suitable for its small form factor but has a lower resolution of ADC (12-bit) and higher noise floor (${\approx}$8dB), compared to off-the-shelf Wi-Fi access points, typically between 2-4dB. The B205-mini is powered by the UAV battery with the additional accessories mounted on the chassis as shown in figure \ref{fig:drone}. The total additional payload  is approximately 5oz. and achieved an average flight time of 15 minutes.

\subsection{UAV positioning system}

Coordinates of the UAV, $pos_k$ is obtained by the Real-time kinematic (RTK) positioning system. RTK is used to enhance the precision of the fix derived from global navigation and satellite systems like GPS, GLONASS, etc. It combines  the phase of the GNSS carrier, its information content and the signal from a reference station to provide real-time correction that yields centimeter level accuracy of the fix. 
With the advent of low cost RTK receivers like CUAV C-RTK GPS, Drotek XL RTK GPS, etc.
supported by open source community~\cite{px4}, it has become essential for precise navigation of UAVs. Although many off-the-shelf modules exist, SparkFun GPS-RTK Board~\cite{sparkfun} based on u-blox NEO-M8P-2~\cite{u-blox} module is best suited for its small form factor and ease of use. This unit reports GPS location with a precision of 1cm and accuracy of 2.5cm, but may vary as seen in Figure~\ref{fig:drift}. Figure~\ref{fig:rtk_base} shows the RTK-GPS setup used as the base connected to a laptop. A similar setup acts as rover, connected to on-board Raspberry Pi.

Figure~\ref{fig:rtk_block} shows the block diagram of the base and the rover units in our setup. The base is the static unit, which transmits phase correction stream over a direct Wi-Fi link to the rover mounted on the UAV. The base unit is configured to transmit RTCM (Radio Technical Commission for Maritime Services) v3 messages that the rover forwards to the RTK-GPS unit, which calculates high precision location. We use a laptop for the base and a Raspberry Pi as the rover due to its small size and powered it with an on-board battery. The Raspberry Pi stores the coordinates and the GPS time as a tuple during the flight, which is used for DoA calculation.  

\note{
\subsection{Total Time for Localization}
The time taken by {\abbrev} to determine the location of the source is determined by: 1) the time for signal acquisition, and 2) the processing time. However, the processing time is negligible compared to the signal acquisition time, due to the logarithmic run time complexity of {\abbrev} in Section 7.  
The time for signal acquisition is determined by: a) the time required to capture the signal at each position, and b) the traversal time of the UAV. 
The duration of a Non-High Throughput (Non-HT) legacy Wi-Fi short preamble is only $8\mu$s, while the duration of a standard preamble in LoRa transmission with a spreading factor of 7 is $10$ms \cite{LoRa_duration}.
Therefore, for the experimentation, an effective packet capture time of $10$ms suffices for each position of the UAV to capture either Wi-Fi or LoRa preambles. 
The time required to capture $N$ such signal instances for DoA estimation at each position is given by $10N$ ms and is in the order of milli-seconds. 
Thus, the total duration of the signal captured at two locations is $2{\times}10N{=}20N$ ms. 
For example for $N{=}20$, the total duration of the captured signal is only $20{\times}20{=}400$ms. 
The UAV traversal time is determined by the time to travel to each reception position within a sphere of radius $R$ at each location and the time to traverse from one location to another which are separated by a distance $d$, as shown in figure 1. In practice since $d{\gg}R$, the total traversal time is determined by $d$ and for a UAV traversing at a speed of $v$, this time is given by ${\approx}\frac{d}{v}$ and is in the order of seconds.  
Therefore the total time required for localization is ${\approx}\frac{d}{v}$, which represents the theoretical minimum amount of time required by {\abbrev} to locate a source. 
For example, for DoA estimation and localization from two locations separated by $d{=}20$ meters for a UAV traversing at a nominal velocity of $10$m/s \cite{faa_2020}, the total time for localization is ${\approx} \frac{d}{v}{=}2$ seconds. 
}

\section{Evaluation}
\label{sec:results}

\begin{figure*}
  \centering
  \subcaptionbox{SNR of the received signal
  \label{fig:snr}}[.3\textwidth][c]{%
    \includegraphics[width=0.9\linewidth]{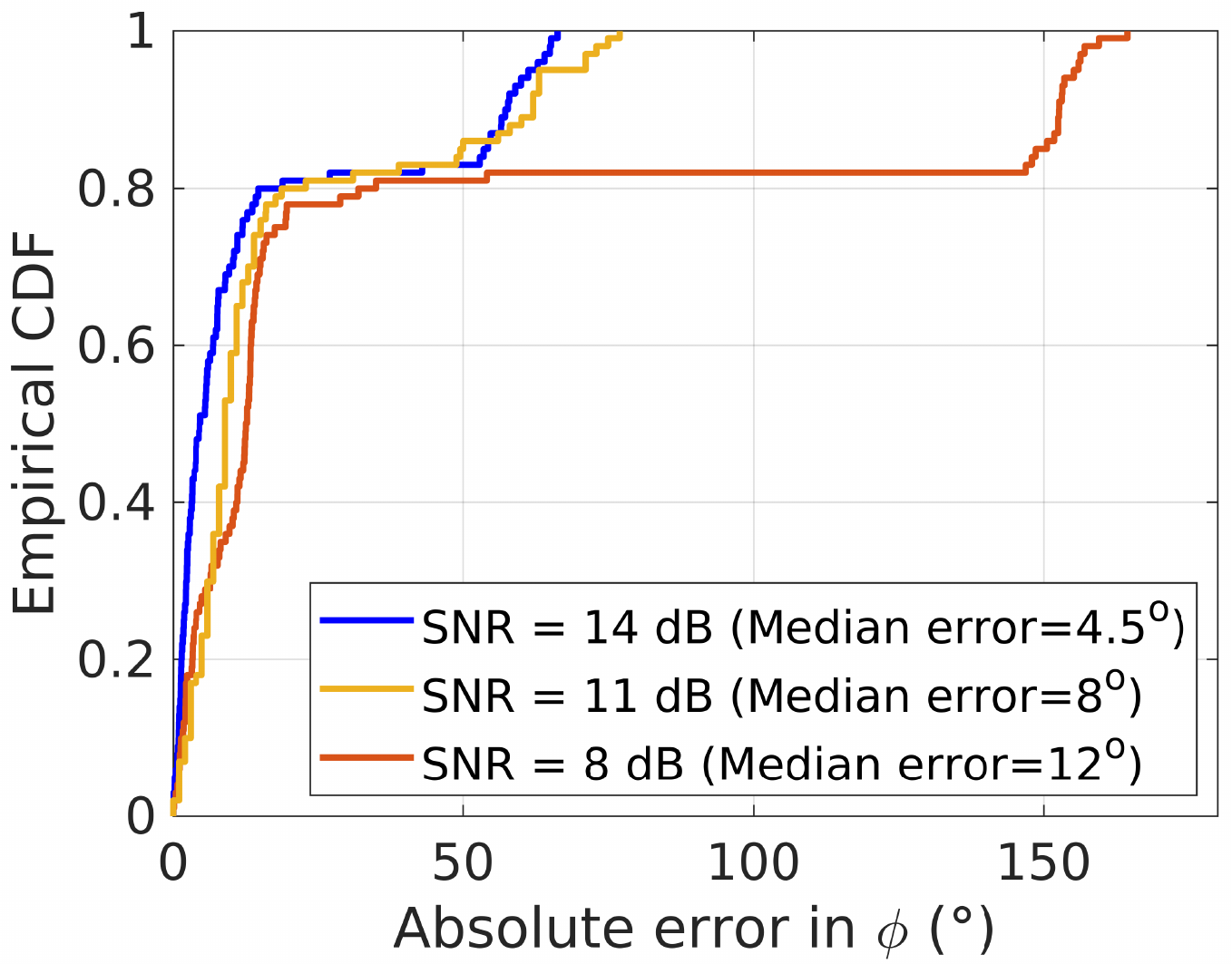}}\quad
  \subcaptionbox{Bandwidth of the Tx signal \label{fig:bw}}[.3\textwidth][c]{%
    \includegraphics[width=0.9\linewidth]{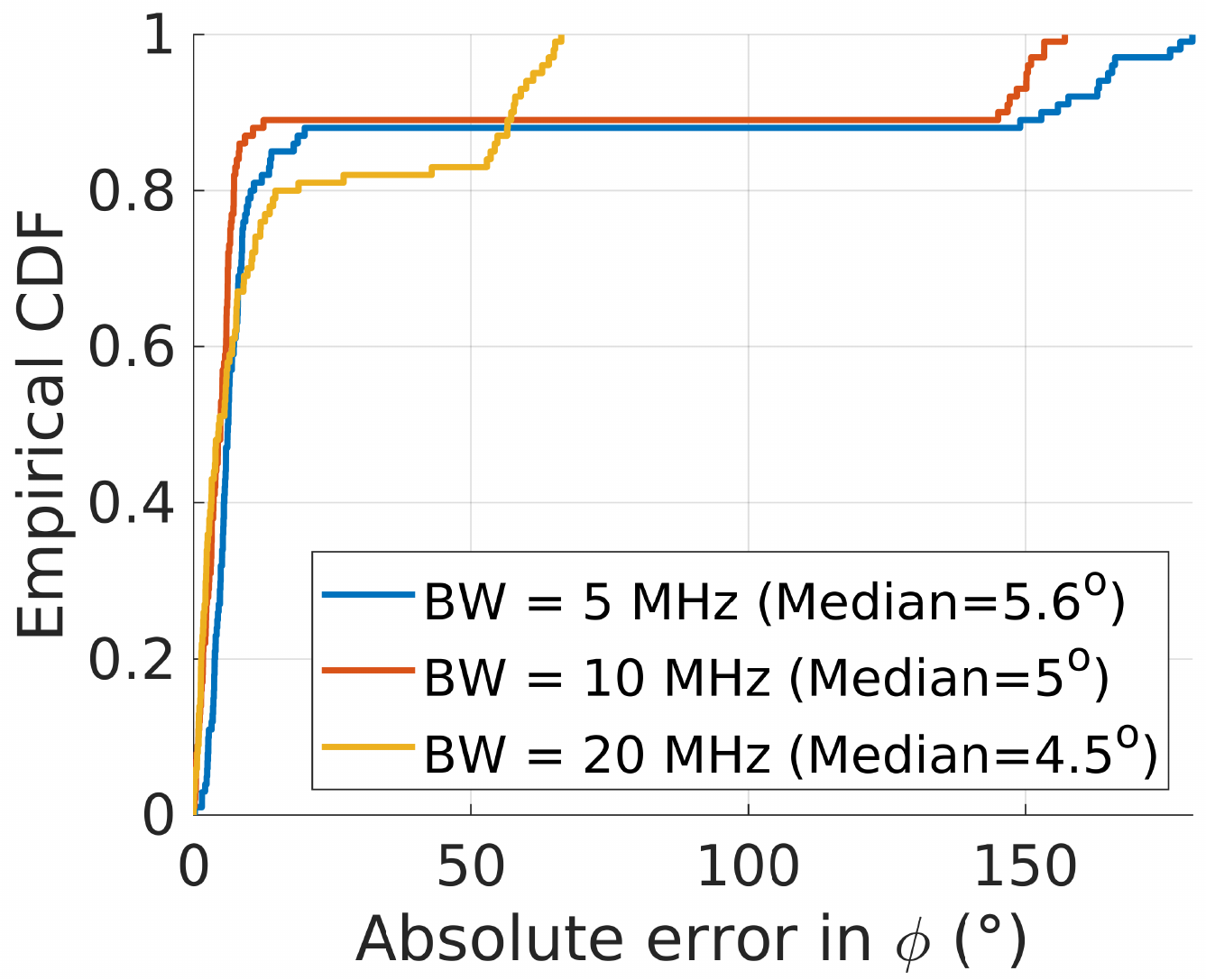}}\quad
  \subcaptionbox{LOS and NLOS scenarios \label{fig:nlos}}[.3\textwidth][c]{%
    \includegraphics[width=0.9\linewidth]{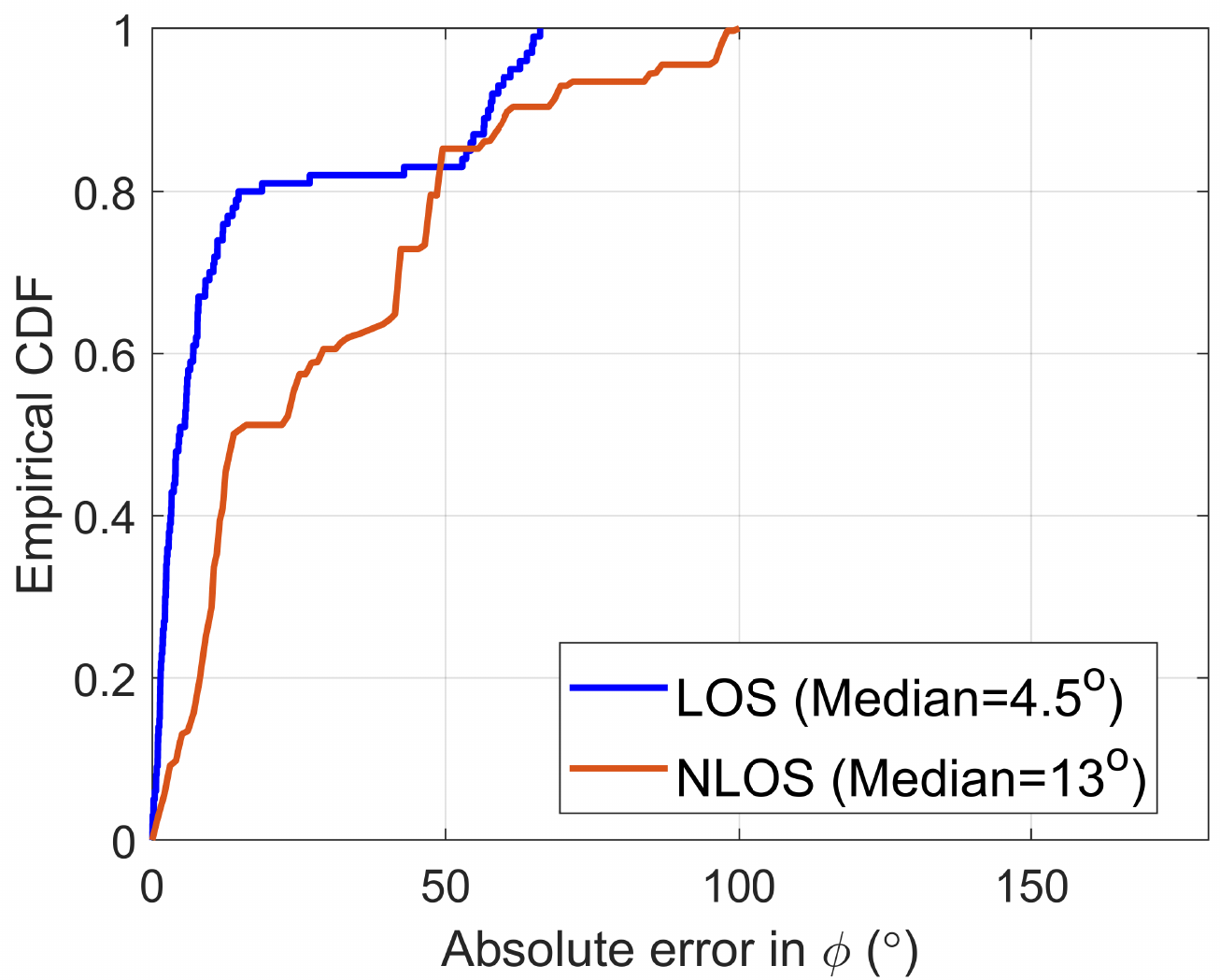}}
  \caption{The impact of external parameters on the DOA accuracy. SNR=14dB, BW = 20MHz
  }
  \label{fig:snr_bw_nlos}
\end{figure*}

\begin{figure*}
	\centering
	\begin{subfigure}{0.234\linewidth}
		\includegraphics[width=\linewidth]{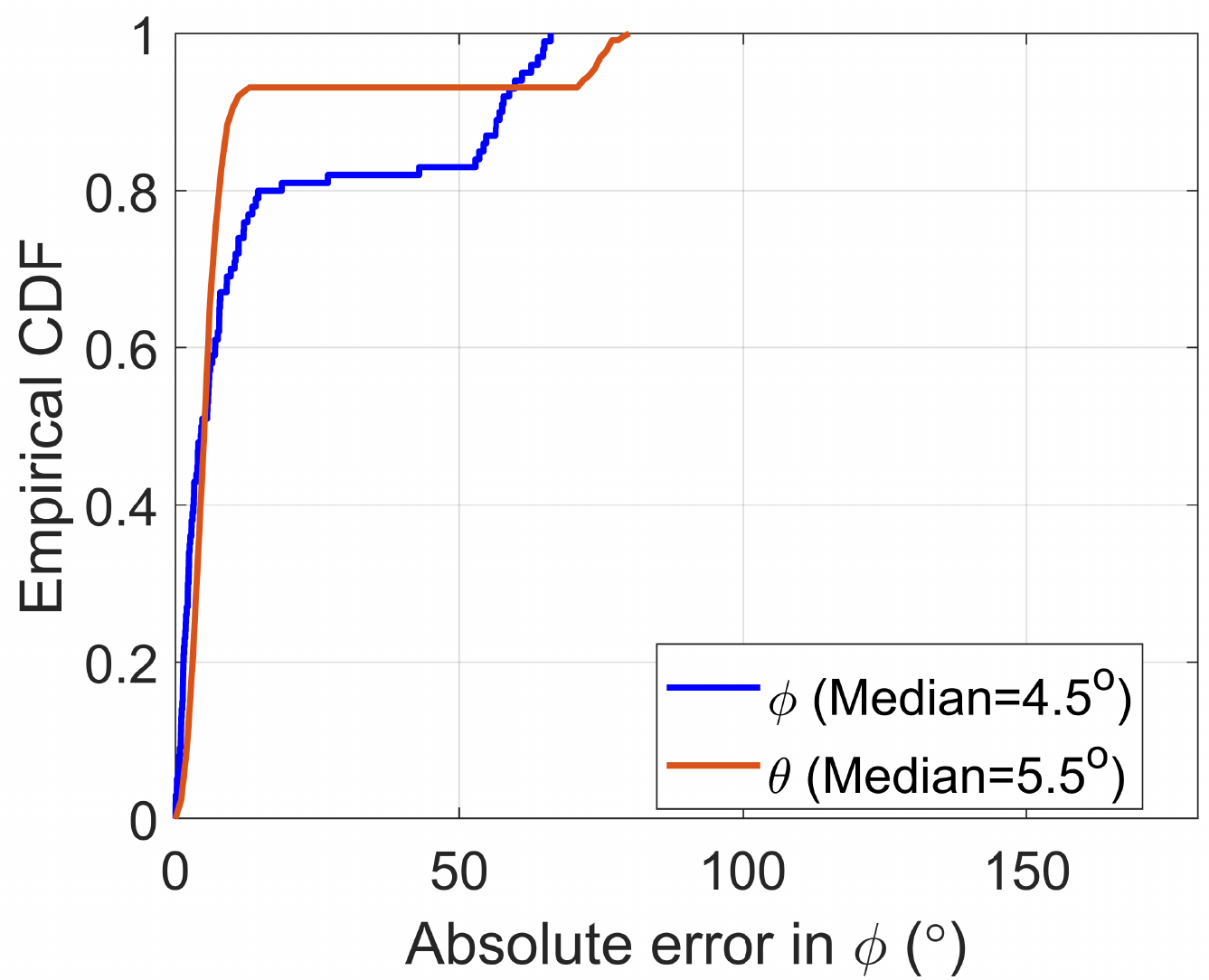}
		\caption{DOA in azimuth, elevation for Wi-Fi}
		\label{fig:doa_error}       
	\end{subfigure}
    \enskip
	\begin{subfigure}{0.234\linewidth}
		\includegraphics[width=\linewidth]{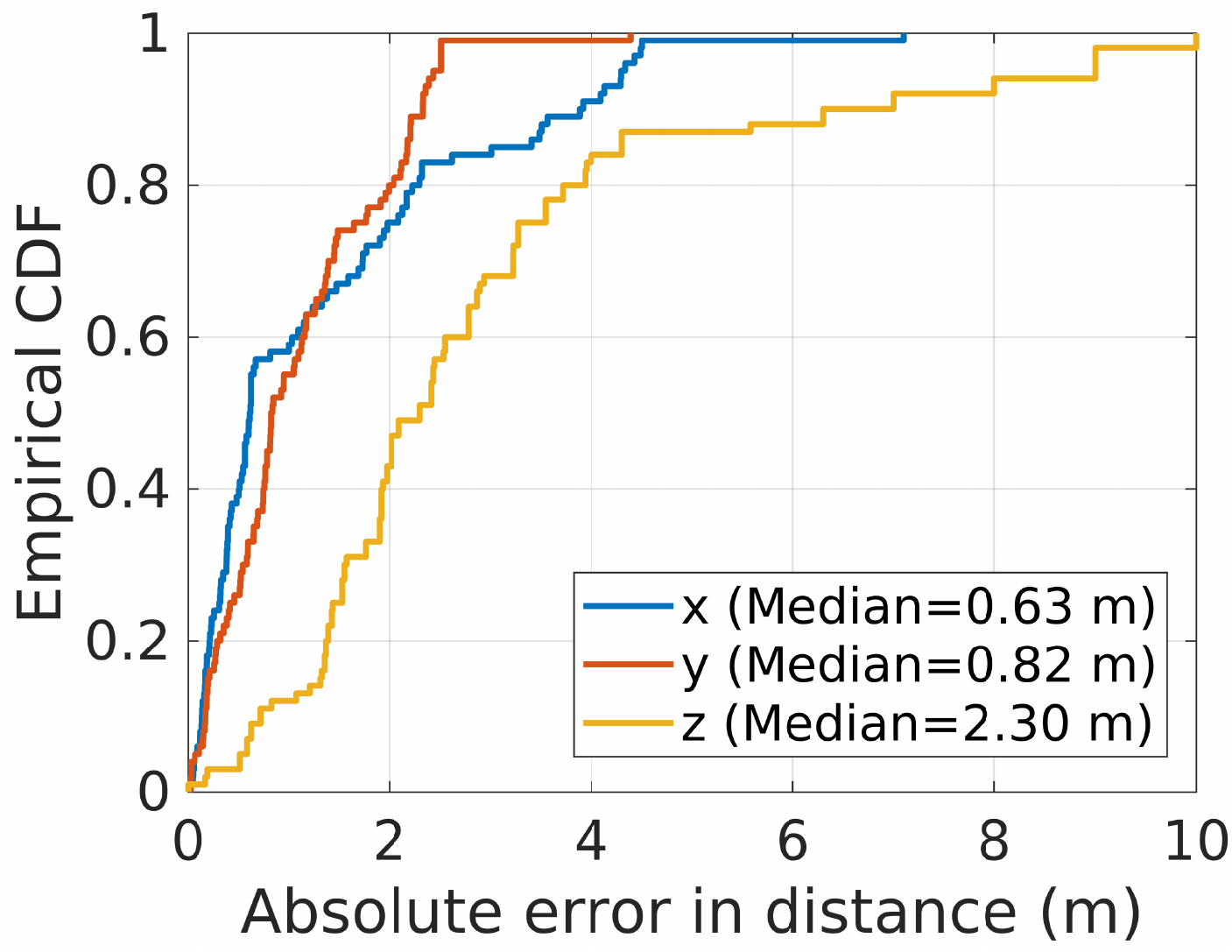}
		\caption{Localization in X, Y and Z directions for Wi-Fi}
		\label{fig:xyz_error}
	\end{subfigure}
    \centering
    \enskip
	\begin{subfigure}{0.234\linewidth}
		\includegraphics[width=\linewidth]{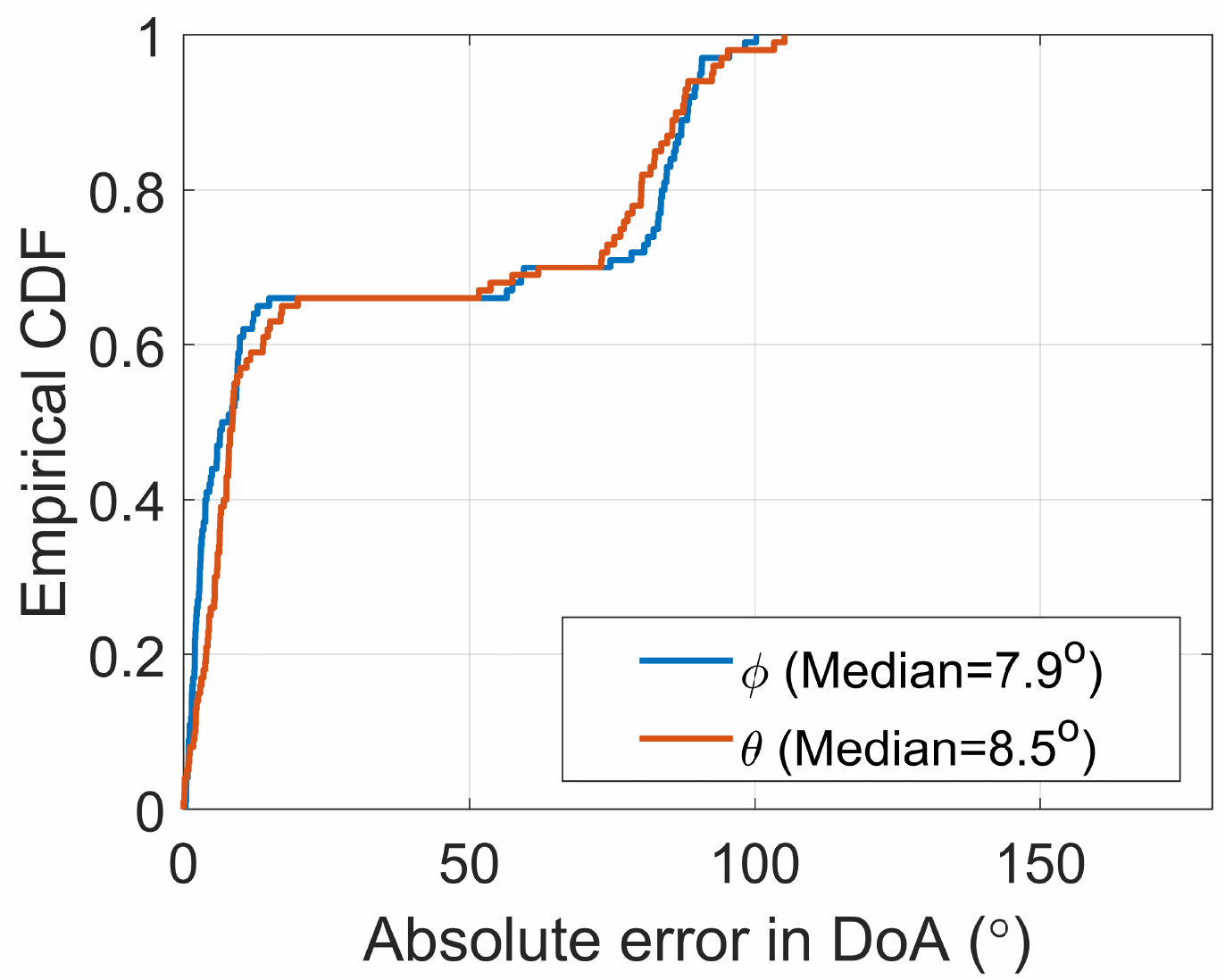}
		\caption{DOA in azimuth, elevation for LoRa}
		\label{fig:doa_lora}       
	\end{subfigure}
    \enskip
	\begin{subfigure}{0.234\linewidth}
		\includegraphics[width=\linewidth]{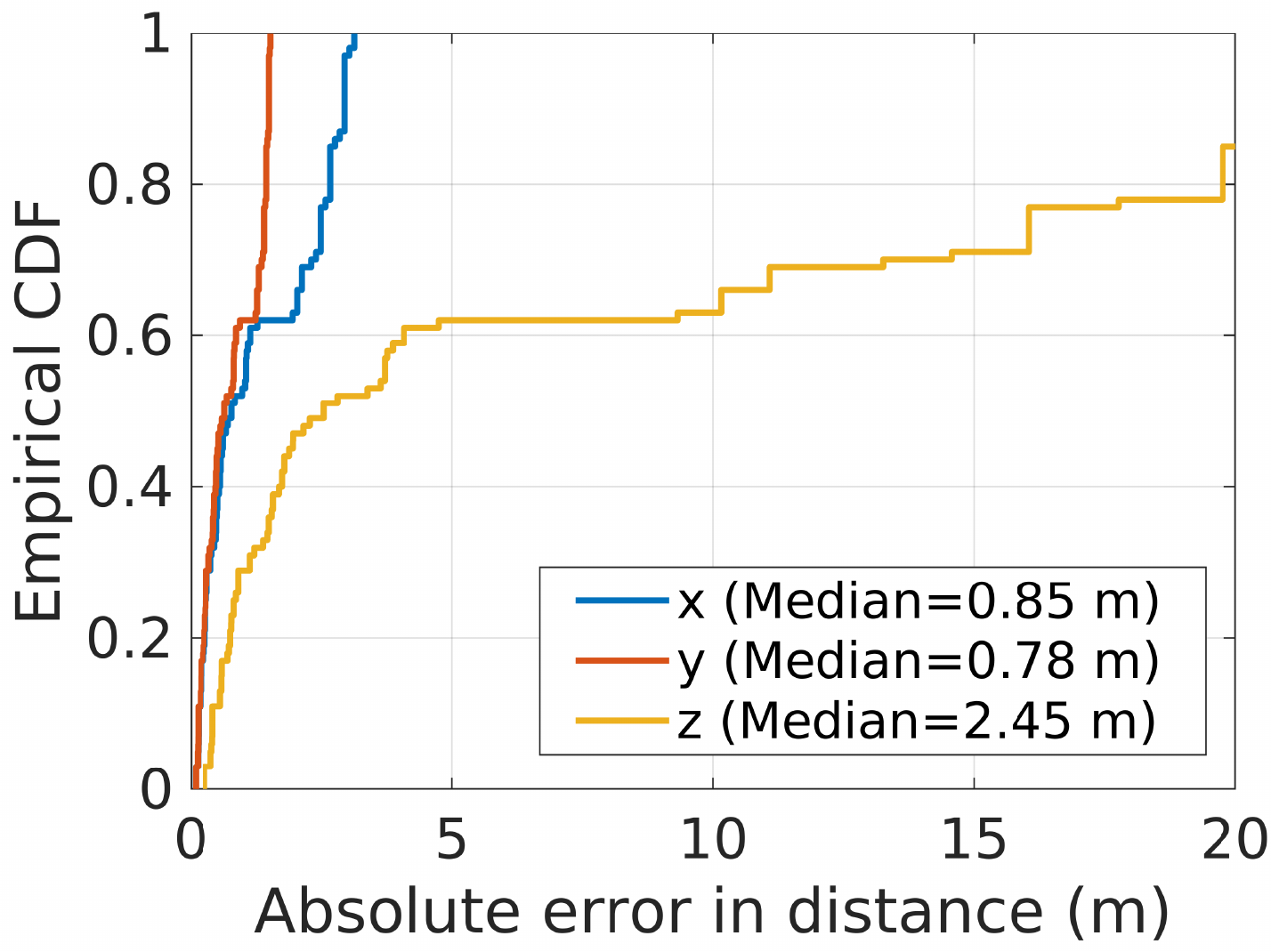}
		\caption{Localization in X, Y and Z directions for LoRa}
		\label{fig:xyz_lora}
	\end{subfigure}
	\caption{The accuracy of {\abbrev} for Wi-Fi [(a) \& (b)] and LoRa signals [(c) \& (d)]} 
	\label{fig:angle}
	\vspace{-10pt}
\end{figure*}

The performance of {\abbrev} is investigated with varying parameters in different scenarios.
We present the results in four categories: 1) Exploring the internal parameter space, 2) Evaluating the external factors, 3) DoA and location accuracy for a Wi-Fi transmitter and 4) Localizing a LoRa transmitter. 
For the evaluation using Wi-Fi signals, we used non-high throughput legacy preamble at 20MHz bandwidth (unless otherwise specified). 
The total number of reception points, $N$ for each location is fixed at 20 for all experiments.

\subsection{Impact of Internal Parameters}
\label{sec:parameter_space}

In this experiment, the UAV hovers 
${\approx}20$m away from the source such that the average received SNR of the signals about 11dB. The purpose of this experiment is to evaluate the role of various parameters in {\abbrev} at moderately low SNR. We discard all the signals captured $\leq$10.5dB and $\geq$11.5dB. 
Figure~\ref{fig:points_pkts_clusters} shows the error in DoA for different tunable parameters of {\abbrev}. 
Figure~\ref{fig:NumPoints} shows the effect of increasing number of reception points $M$ on DoA accuracy 
with a median error of ${<}6^{\circ}$ in azimuth at moderately low SNR with $M$=16. We notice a diminishing return beyond 16 points and therefore can be kept constant for resource constrained UAV.
The effect of the radius of the dominant cluster on the accuracy of the DoA is shown in Figure~\ref{fig:radius}. 
Clearly, the DoA accuracy improves with decreasing radius, 
 which is also used as termination condition in Algorithm \ref{algo:doa}. 
The figure shows a median error of $<6^o$ when the radius is $5^o$ and no significant improvement in the DoA accuracy is observed below this radius. Hence, the threshold for the radius of the dominant cluster in Algorithm \ref{algo:doa} is set to $R_{th}{=}5^o$.
Figure~\ref{fig:NumClusters} shows the improvement in DoA accuracy with the number of iterations $n$ in Algorithm \ref{algo:doa}, when $M$ varies from $\ceil[\big]{N/2}{=}10$ to $N{-}2{=}18$.
This indicates that $\mathcal{N}$ varies from ${20 \choose 10}{=}184,756$ to ${20 \choose 18}{=}190$. 
Results show that the median error in DoA reduces from 45$^{\circ}$ without any clustering to ${<}$5$^{\circ}$ by clustering the DoA from 40 random instances of $\mathcal{F}_n$ for each $M$. A diminishing return is observed beyond 40 
and hence 
$\mathcal{N}_{max}$ in Algorithm \ref{algo:doa} for each $M$ is set to 40.

\subsection{Impact of the External Factors}
\label{sec:external_factors}

Here, we evaluate {\abbrev} with various parameters 
that are external to the system like Bandwidth, SNR and NLOS. These are properties of the signal that cannot be controlled and figure~\ref{fig:snr_bw_nlos} shows the performance with these parameters. For these experiments $M$ varies from 10 to 18 and $R_{th}{=}5^{\circ}$.
Figure~\ref{fig:snr} shows that {\abbrev} achieves a median accuracy of 4.5$^{\circ}$, 8$^{\circ}$ and 12$^{\circ}$ at SNR of 14dB, 11dB and 8dB respectively with line of sight to the transmitter. 
Although there is a higher error at low SNR, we believe that it provides a wider direction for the UAV, which can be used to fly towards the target to improve the SNR and the DoA accuracy.
Generally, wider signal bandwidth leads to higher accuracy in temporal alignment and DoA. {\abbrev} is evaluated at 14dB SNR and line-of-sight for three different bandwidths: 5, 10 and 20MHz, which are also part of the IEEE 802.11a standard~\cite{802_11_spec}. Figure \ref{fig:bw} shows a median DoA accuracy of 4.5$^{\circ}$ for a 20MHz signal and 5.6$^{\circ}$ for 5MHz signal. This also shows that the effect of bandwidth of the signal on the DoA accuracy is small. 
\note{Moreover, the 75\textsuperscript{th} percentile error in DoA estimates are 9$^{\circ}$, 13$^{\circ}$ and 14.5$^{\circ}$ at SNRs of 14dB, 11dB and 8dB respectively as seen in figure~\ref{fig:snr}. 
In figure~\ref{fig:bw}, the 75\textsuperscript{th} percentile error in DoA is 9$^{\circ}$ for a 20MHz signal and 7.2$^{\circ}$ for 5MHz signal. These results corroborate the reliability of the DoA estimates of {\abbrev} and the robustness to the external factors in outdoor environments.} 

Localization in outdoor is often presented with non-line-of-sight (NLOS) scenarios. We compared the performance of {\abbrev} in NLOS by flying the UAV behind large evergreen trees that provides a strong shadowing effect. We notice a drop in SNR of 5dB in NLOS compared to LOS for the same distance between the target and UAV. Figure~\ref{fig:nlos} shows the error in angle estimation in NLOS compared to LOS. 
The median error of 13$^{\circ}$ in a highly blocked low SNR  demonstrates that {\abbrev} is efficient in outdoor scenarios.

\subsection{Accuracy of Wi-Fi Localization}





\note{
The accuracy of the azimuth and elevation angles and the location estimates are evaluated at an SNR of 14dB with 20 reception points and a threshold radius of 5$^{\circ}$, across 1000 experimental trials. Figure~\ref{fig:doa_error} shows a median error of 4.5$^{\circ}$ and 5.5$^{\circ}$ and a 75\textsuperscript{th} percentile error of $9^{\circ}$ and $7^{\circ}$ for azimuth and elevation respectively.   
Figure~\ref{fig:xyz_error} shows a median accuracy of 0.63m, 0.82m and 2.3m in X, Y and Z directions. Position error is calculated as $\sqrt{(\Delta_x^2{+}\Delta_y^2)}$, which indicates a median error of 1.03m in 2D, and as $\sqrt{(\Delta_x^2{+}\Delta_y^2{+}\Delta_z^2)}$ indicating an error of 2.5m in 3D. 
The figure also demonstrates a 75\textsuperscript{th} percentile localization accuracy of 1.8m, 1.4m and 3.1m in X, Y and Z directions, and consequently the location accuracy is ${\leqslant}2.2$m in 75\% of the trials. These results show the high reliability of the location estimates of a 20MHz signal from {\abbrev} compared to other source-localization literature shown in Table~\ref{tab:comparison}.
}


\subsection{Accuracy of LoRa Localization}
\label{sec:lora}

We also demonstrate the generality of {\abbrev} 
using a LoRa transmitter as an example. We use an Adafruit Feather M0 LoRa module \cite{feather_m0} to periodically transmit a LoRa packet with the standard preamble at a frequency of 915MHz, bandwidth of 125KHz and a spreading factor of 7. 
The same UAV setup is used to capture the LoRa signals at 20 reception points. 
Figure \ref{fig:doa_lora} shows a median error of 7.9$^{\circ}$ and 8.5$^{\circ}$ in the azimuth and elevation angles respectively at an SNR of 30dB. Figure~\ref{fig:xyz_lora} shows the error in the transmitter localization in x, y and z directions.
{\abbrev} achieves a median accuracy of 0.85m, 0.78m and 2.45m in x, y and z directions, which indicates a median position error of 1.15m in 2D and 2.70m in 3D. 
This shows that {\abbrev} can accurately localize an RF source transmitting any generic waveform with a repetitive pattern. 

\begin{figure}
\centering
        \begin{subfigure}[b]{0.24\textwidth}
		\includegraphics[width=\textwidth]{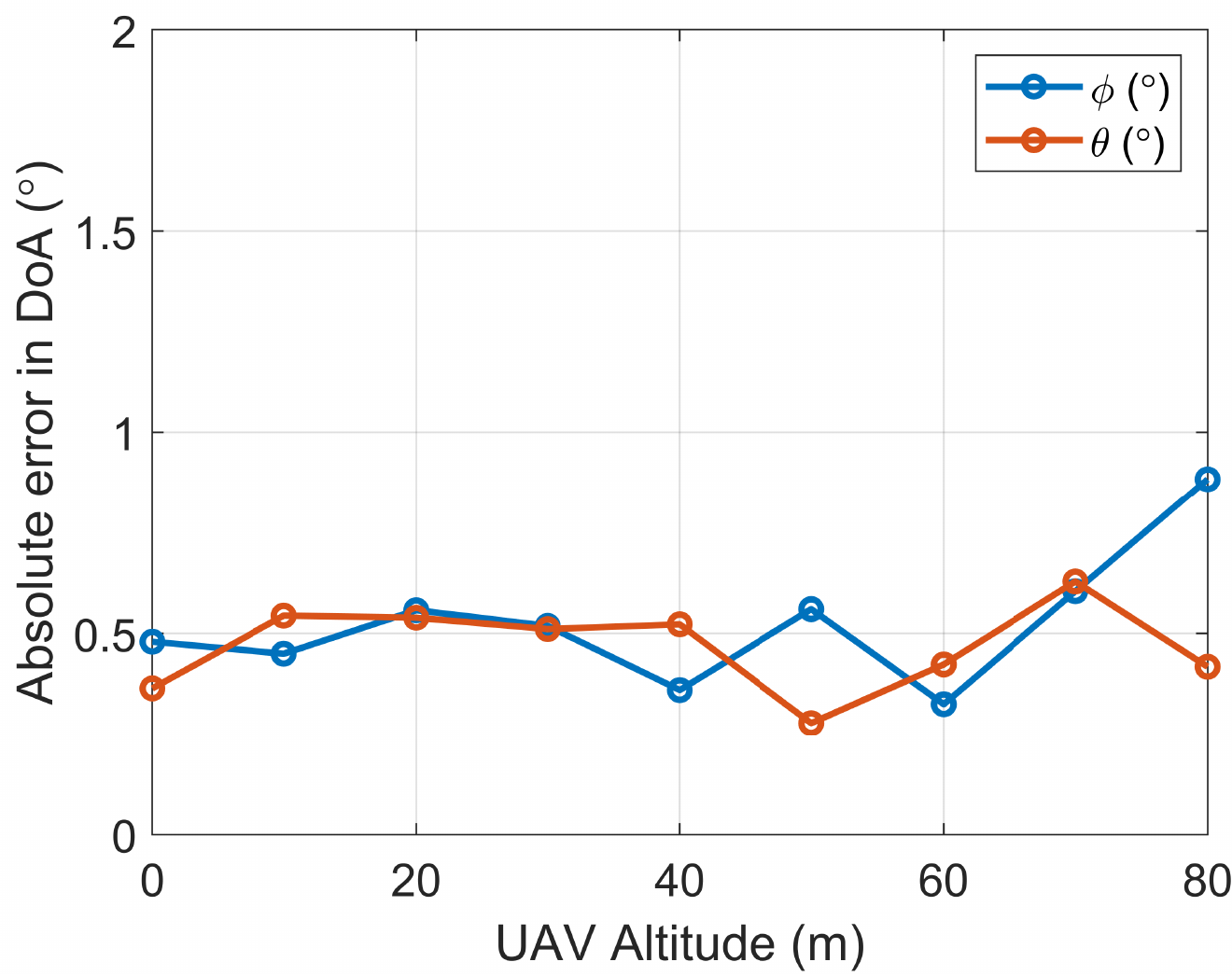}
                \caption{DoA accuracy at various UAV altitudes}
                \label{fig:doa_height}
        \end{subfigure}
        \begin{subfigure}[b]{0.24\textwidth}
                \includegraphics[width=\textwidth]{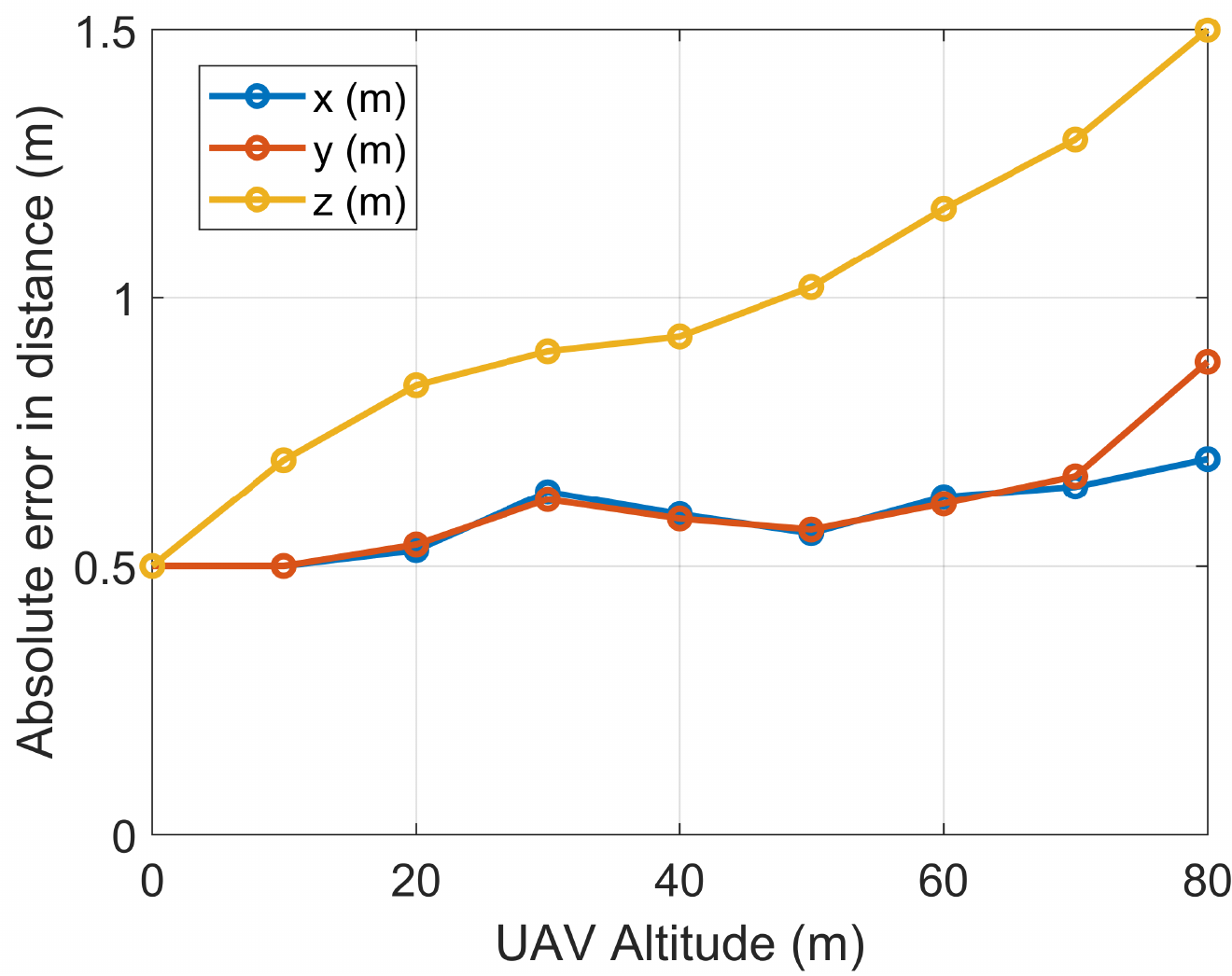}
                \caption{Localization accuracy at various UAV altitudes}
                \label{fig:localization_height}
        \end{subfigure}
\caption{Accuracy of {\abbrev} with UAV altitude.}
\label{fig:height}
\vspace{-10pt}
\end{figure}

\note{
\subsection{Impact of UAV Altitude}
\label{sec:altitude}
The experiments were conducted at relatively low altitudes due to the following practical constraints: 
a) restrictions imposed by the Federal Aviation Administration (FAA) mandate the maximum altitude for experiments conducted with small UAVs, which are further limited at locales close to airports~\cite{faa_2020}, and 
b) the low transmit power of the source (USRP B210) and high noise floor of the receiver (USRP B205) also limit the maximum altitude at which the signal can be decoded for localization. 
Therefore, the performance of {\abbrev} is investigated at varying altitudes of the UAV using a robust simulation environment.
For the evaluation, a Wi-Fi source emitting a non-high throughput legacy preamble at 20MHz bandwidth using an omnidirectional antenna is employed as the transmitter. 
The UAV flies to two random locations at various altitudes and hovers within a sphere of radius 1 meter, collecting signals using one omnidirectional antenna at various positions within the sphere. 
The errors in the UAV position measurement are collectively modeled as a random variable $pos_k{\in}{\sim} \mathcal{N}(\mu,\sigma^2)$ for all $k{\in}N$, with $\mu{=}0$ and $\sigma^2{=}2.5$cm \cite{sparkfun} and a two-ray multipath propagation model \cite{two_ray_model} is used for the source-to-UAV channel. 
Figure \ref{fig:height} shows the median error in DoA estimation and localization at various altitudes of the UAV, by averaging over all SNR (5-15dB) and angles of the transmitter ($\phi_{tx},\theta_{tx}{\in}[-\pi,\pi]$).
A consistently low median error of ${<}1^{\circ}$ is observed in both azimuth and elevation angle estimates over all UAV altitudes in figure \ref{fig:doa_height}, demonstrating the robustness of DoA estimation using {\abbrev}.
Figure~\ref{fig:localization_height} shows that the median localization error in both X and Y directions is $0.5{-}0.85$m and exhibits minimal variation with the altitude of the UAV. 
The median error in Z direction marginally increases with increasing altitude of the UAV due to: a) the lower average signal-to-noise ratio at larger altitudes, and b) the inaccuracies in the elevation angle estimates are extrapolated at higher altitudes, i.e. the higher the altitude, the more the elevation angle inaccuracies perturb the Z location.
However, we observe that the error in Z location is at most $1.5$m even at an UAV altitude of 80m. 
}

\section{Conclusion}
\label{sec:conclusion}

This work presents {\abbrev}, an outdoor localization system using UAV with limited hardware and processing capacity. We leverage the mobility of the UAV to asynchronously capture signals from multiple positions, detect and extract a signature and align those in time to emulate signal reception from a synchronous array to calculate highly accurate DoA. We leverage combinations of signals to expand the candidate DoAs and perform spatial clustering to minimize error. The proposed algorithm has low computational complexity and is robust to various internal and external sources of errors. Finally, we implement {\abbrev} on commodity hardware and demonstrate a median error in location of 1.09m in 2D and 2.6m in 3D. Our results also show the generality by locating both LoRa and Wi-Fi signals while immune to factors like wind and position errors. 
Lastly, we acknowledge that leveraging higher dimensionality of the signal space to improve accuracy and locating mobile emitters are both worthwhile extensions to pursue in future as it requires rethinking the mathematical and signal processing apparatus of {\abbrev}.

\IEEEoverridecommandlockouts.
\vskip -1\baselineskip plus -1fil
\begin{IEEEbiography}[{\includegraphics[width=1in,height=1.25in,keepaspectratio]{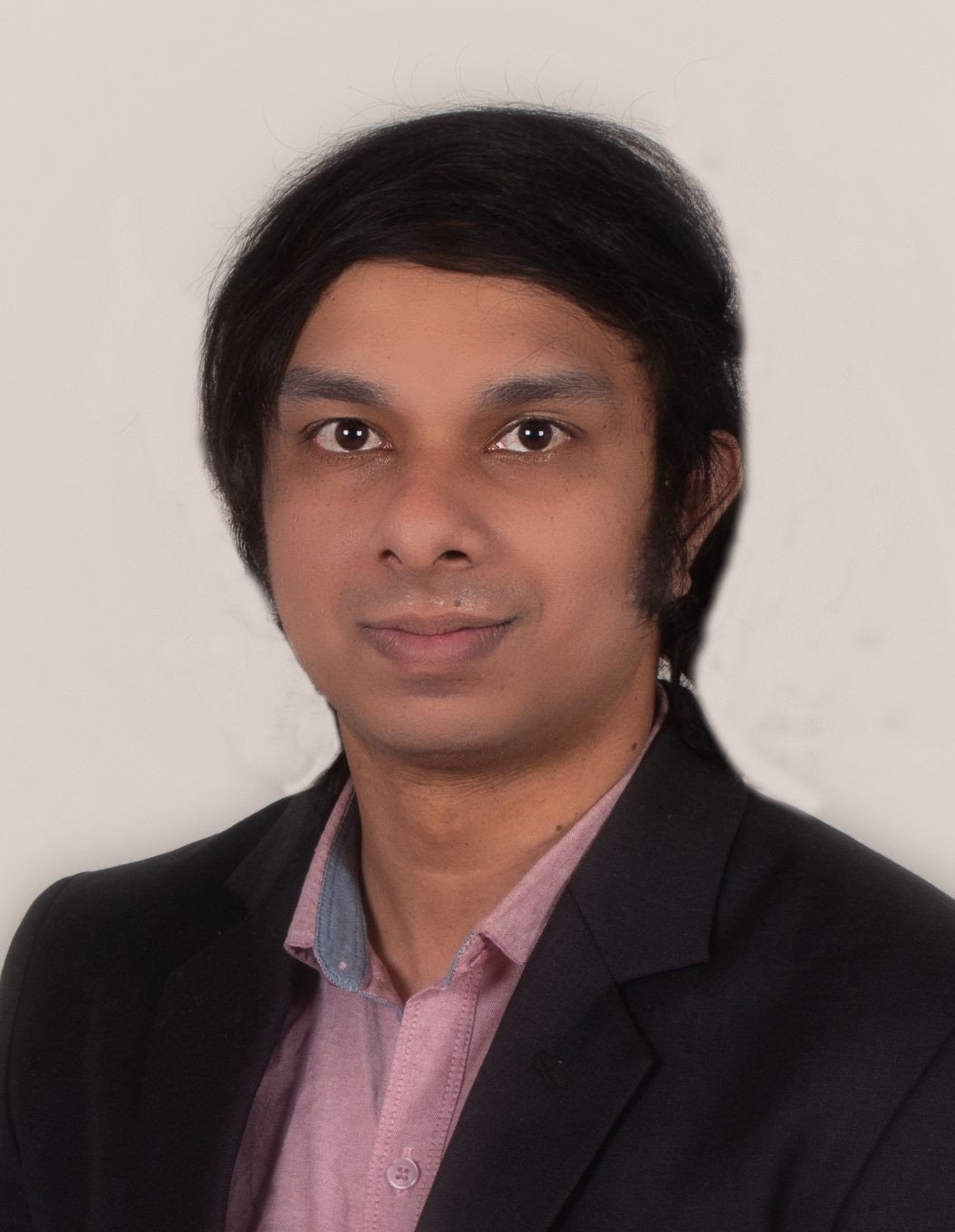}}]%
{Maqsood Ahamed Abdul Careem} received the MS.c degree in Electrical \& Computer Engineering from University at Albany and the BS.c degree in Electrical and Electronic engineering from the University of Peradeniya, Sri Lanka, in 2014. He is currently pursuing the Ph.D. degree with the Department of Electrical and Computer Engineering, University at Albany SUNY, Albany, NY, USA, where he is a member of the Mobile Emerging Systems and Applications (MESA) Lab. His research interests include learning based optimization of wireless systems, heterogeneous wireless networks, spectrum management and distributed wireless systems.
\end{IEEEbiography}
\vskip -2\baselineskip plus -1fil
\begin{IEEEbiography}[{\includegraphics[width=1in,height=1.25in,keepaspectratio]{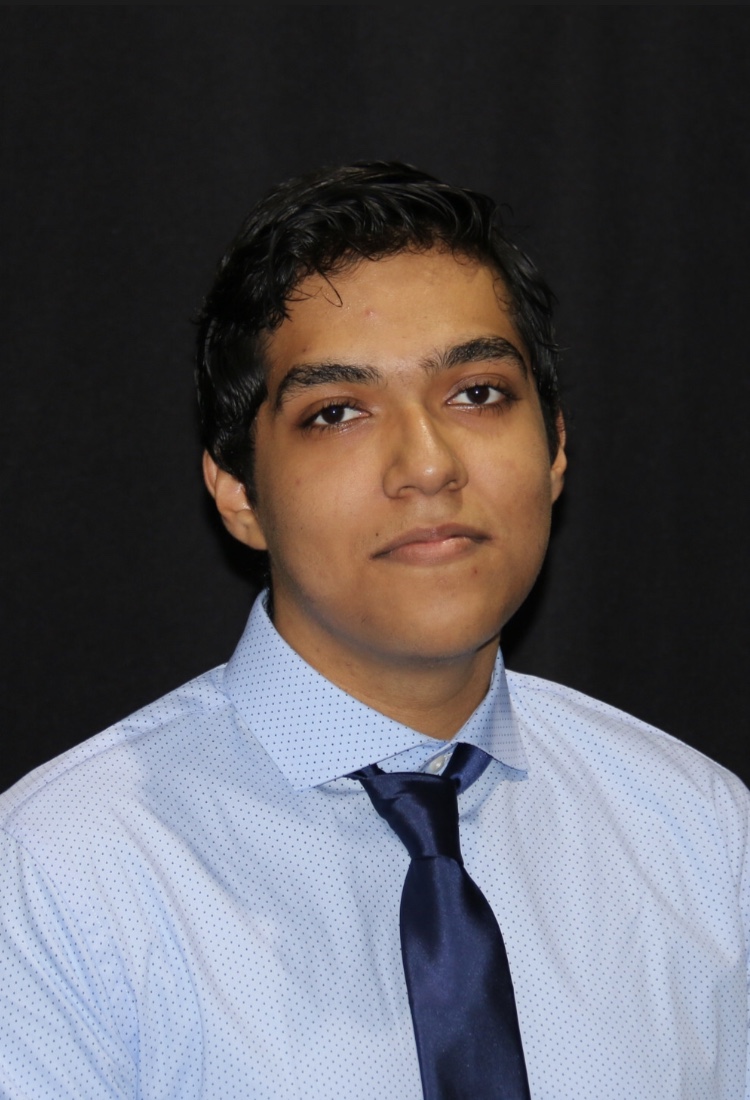}}]%
{Jorge Gomez} is a Firmware Engineer and is currently pursuing his Master's degree at the NYU Tandon School of Engineering. He received his Bachelor's from the University at Albany, SUNY. His interests include wireless communications applications and fitness applications in embedded systems.
\end{IEEEbiography}
\vskip -1\baselineskip plus -1fil
\begin{IEEEbiography}[{\includegraphics[width=1in,height=1.25in,keepaspectratio]{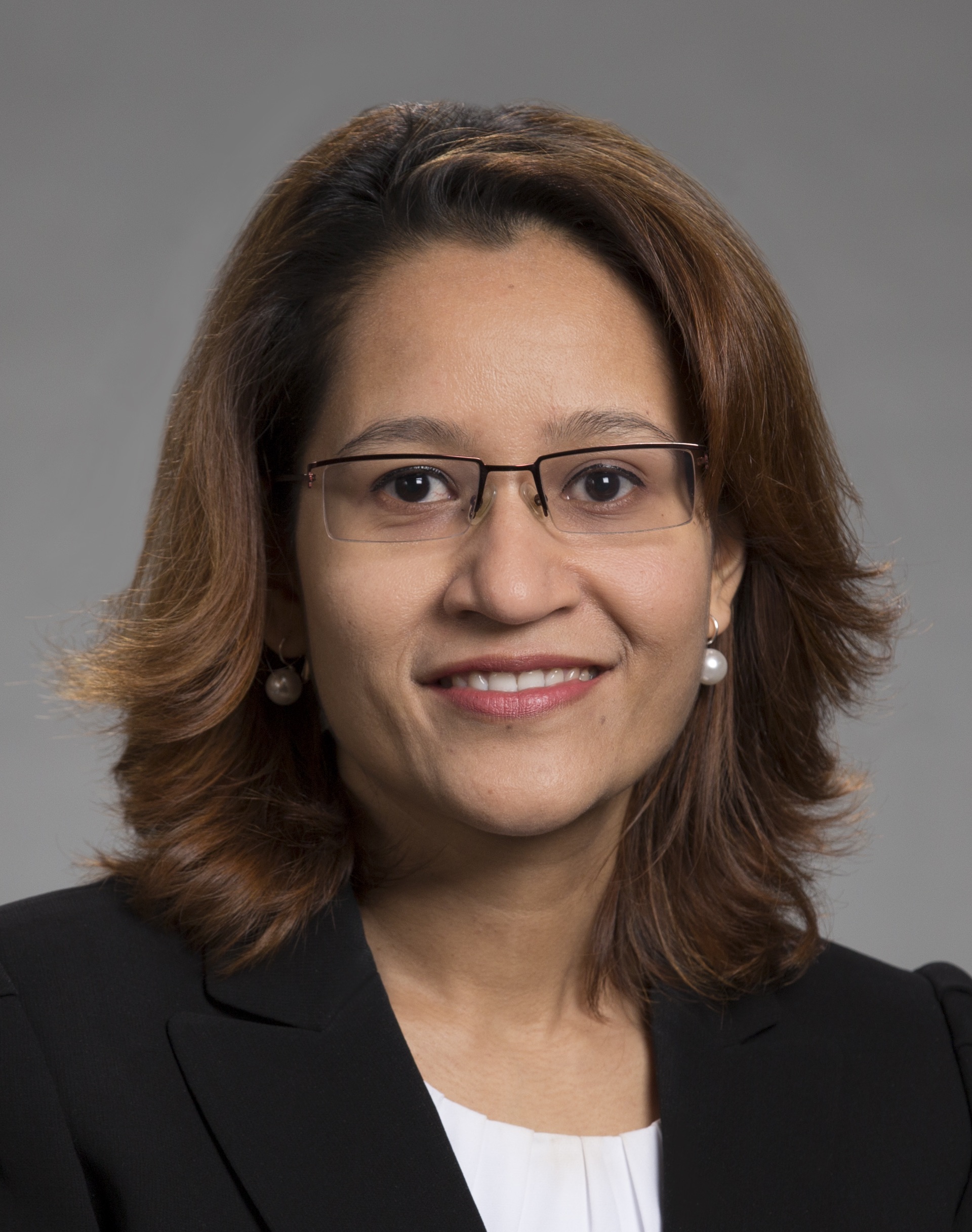}}]%
{Dola Saha} is an Assistant Professor in the Department of Electrical \& Computer Engineering at the University at Albany, SUNY. She co-directs the Mobile Emerging Systems and Applications (MESA) Lab at UAlbany. Prior to that, she was a Research Assistant Professor 
at Rutgers University. Before that, she was a Researcher at NEC Laboratories America. She received her PhD in Computer Science from the University of Colorado Boulder. She is also a recipient of the Google Anita Borg scholarship for her outreach services and academic credentials. Her research interests lie in the crossroads of wireless communication, signal processing and machine learning applications.
\end{IEEEbiography}
\vskip -1\baselineskip plus -1fil
\begin{IEEEbiography}[{\includegraphics[width=1in,height=1.25in,keepaspectratio]{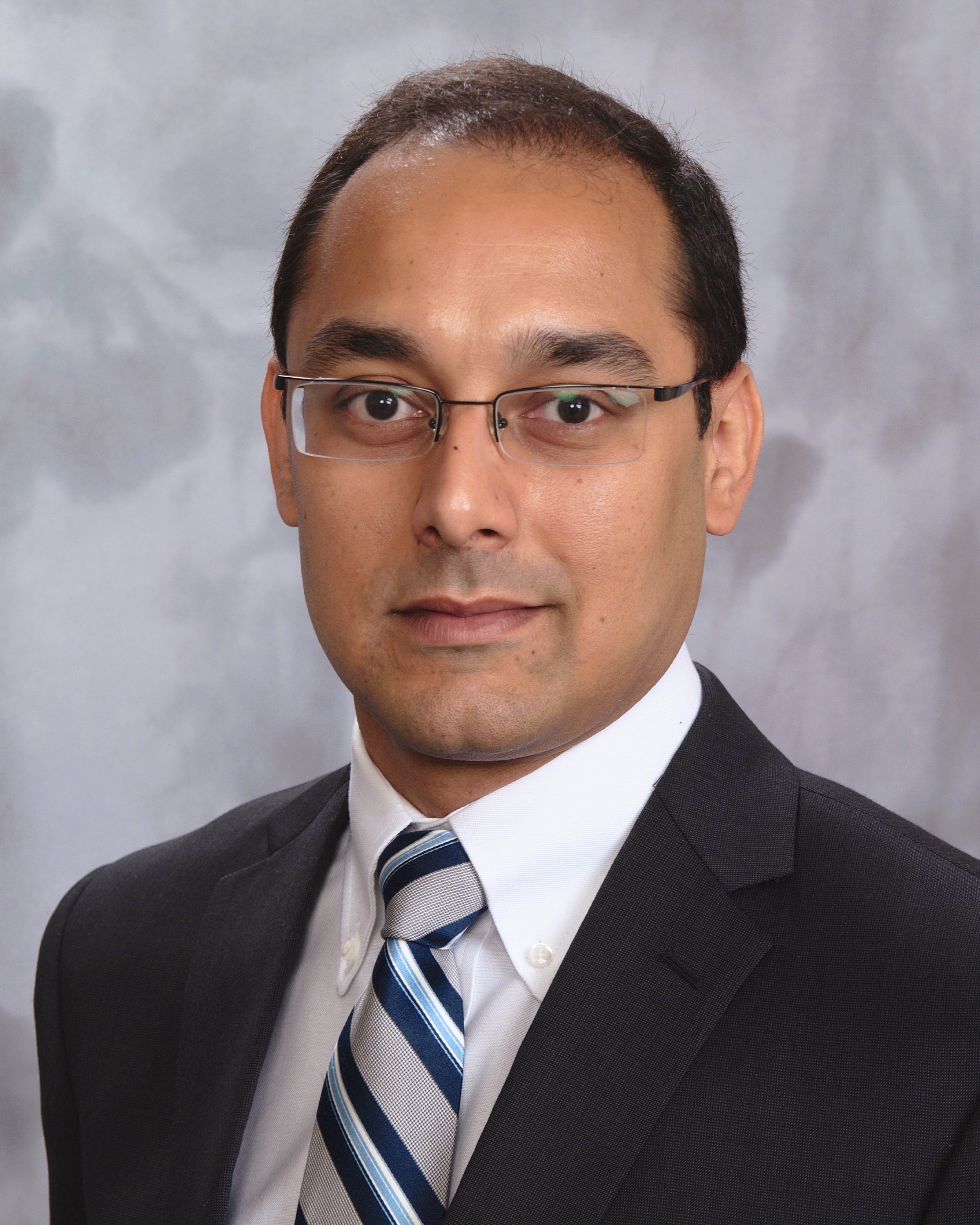}}]%
{Aveek Dutta} is an Assistant Professor with the Department of Electrical and Computer Engineering, University at Albany SUNY. He received his M.S. and Ph.D. degrees in electrical engineering from the University of Colorado, Boulder, CO, USA and Postdoctoral experience at Princeton University. He co-directs the MESA lab at University at Albany, which primarily investigates open problems in emerging heterogeneous wireless networks with an emphasis on system implementation. He has also architected flexible radio platforms and has worked on knowledge representation methods for complex physical layers in cognitive radio networks. His research strives to shape practical systems with strong theoretical foundation.
\end{IEEEbiography}






\end{document}